\documentclass[12pt]{article}

\usepackage{epsf,epsfig}
\usepackage{color,colortbl}
\usepackage{graphics}
\usepackage{axodraw}

\renewcommand{\arraystretch}{1.2}

\def\slash#1{#1 \hskip-0.45em /}
\def\Slash#1{#1 \hskip-0.59em /}
\def\simgt{\rlap{\lower 3.5 pt \hbox{$\mathchar \sim$}} \raise 1pt \hbox {$>$}}
\def\simlt{\rlap{\lower 3.5 pt \hbox{$\mathchar \sim$}} \raise 1pt \hbox {$<$}}

\def\be{\begin{equation}}
\def\ee{\end{equation}}
\def\beq{\begin{eqnarray}}
\def\eeq{\end{eqnarray}}
\newcommand{\bea}{\begin{eqnarray}}
\newcommand{\eea}{\end{eqnarray}}
\newcommand{\beas}{\begin{eqnarray*}}
\newcommand{\eeas}{\end{eqnarray*}}

\newcommand{\ub}{\bar u}
\newcommand{\vb}{\bar v}
\newcommand{\wb}{\bar w}
\newcommand{\LL}{\mathrm{Li}_2}
\newcommand{\cH}{{\cal H}}

\newcommand{\npslash}{{n \!\!\! /}_+ }
\newcommand{\nmslash}{{n \!\!\! /}_- }
\newcommand{\nph}{{\frac{\npslash}{2}}}
\newcommand{\nmh}{{\frac{\nmslash}{2}}}

\newcommand{\xib}{{\bar \xi}}

\newcommand{\hc}{{\mathrm{hc}}}

\newcommand{\Wcp}{W_{c2}}
\newcommand{\gp}{\gamma_{\perp}}

\newcommand{\OI}{{O^{\mathrm{I}}}}
\newcommand{\OII}{{O^{\mathrm{II}}}}

\newcommand{\sceti}{{SCET$_{\mathrm{I}}$}}

\newcommand{\hf}{{\frac{1}{2}}}
\newcommand{\sotimes}{{\,\tilde \otimes\,}}
\newcommand{\eff}{{\mathrm{eff}}}

\newcommand{\eq}[1]{(\ref{#1})}

\begin{document}

\begin{titlepage}

\begin{flushright}
{\small
PITHA~05/21\\
December 23, 2005}
\end{flushright}

\vspace{0.7cm}
\begin{center}
\Large\bf\boldmath
Spectator scattering at NLO in non-leptonic\\ $B$ decays: Tree amplitudes
\unboldmath
\end{center}

\vspace{0.8cm}
\begin{center}
{\sc M.~Beneke and S.~J\"ager}\\
\vspace{0.7cm}
{\sl Institut f\"ur Theoretische Physik E, RWTH Aachen\\
D--52056 Aachen, Germany}
\end{center}

\vspace{1.3cm}
\begin{abstract}
\vspace{0.2cm}\noindent
We compute the 1-loop ($\alpha_s^2$) correction to hard 
spectator scattering in non-leptonic $B$ decay tree amplitudes. 
This forms part of the NNLO contribution to the QCD factorization 
formula for hadronic $B$ decays, and introduces a new rescattering 
phase that corrects the leading-order result for 
direct CP asymmetries. Among the technical issues, we discuss 
the cancellation of infrared divergences, and the treatment 
of evanescent four-quark operators.
The infrared finiteness of our result establishes factorization of 
spectator scattering at the 1-loop order. Depending on the values 
of hadronic input parameters, the new 1-loop correction may have a
significant impact on tree-dominated decays such as $B\to\pi\pi$. 
\end{abstract}

\vfil
\end{titlepage}

\section{Introduction}

The majority of observables at the $B$ factories is connected 
with branching fractions and CP asymmetries of hadronic $B$ decays to 
two charmless mesons, for which strong-interaction effects 
are essential. There is some control over these effects, since 
the decay amplitudes factorize in the heavy-quark limit. In the 
QCD factorization framework~\cite{Beneke:1999br} the matrix 
elements of the effective weak interaction operators take 
the (schematic) expression
\be                    
  \langle M_1 M_2 | Q_i | B \rangle
        = F^{B M_1}(0)\,\, T^\mathrm{I}_{i} * f_{M_2} \phi_{M_2}
          + T^\mathrm{II}_{i} * f_B \phi_{B+} * f_{M_1} \phi_{M_1} *
          f_{M_2} \phi_{M_2}.
\label{eq:qcdfact}
\ee
The long-distance strong-interaction effects are now confined 
to a form factor $F^{B M_1}(0)$ at $q^2=0$, decay constants $f_M$, and 
light-cone distribution amplitudes $\phi_M$. The benefit is that 
information extraneous to two-body $B$ decays is available for 
these, and that the short-distance kernels $T_i^{\rm I,II}$ 
can be expanded in a perturbation series in the strong coupling $\alpha_s$.
Both kernels are currently known from~\cite{Beneke:1999br} 
at order $\alpha_s$. 
While for $T_i^{\rm I}$ this includes a 1-loop correction to 
``naive factorization'', in case of $T_i^{\rm II}$ the 
order $\alpha_s$ contribution is actually the leading term. 
It originates from the tree-level exchange of a 
hard-collinear gluon with the spectator-quark in the 
$B$ meson, as indicated in Figure~\ref{fig2} below. (The 
class of corrections from fermion-loop insertions into the gluon 
propagator is also known~\cite{Burrell:2001pf}. In 
spectator scattering these $n_f$-terms are all connected with the 
hard-collinear scale and make no contribution to the hard 1-loop 
correction, which we compute here.)

In this paper we shall compute the 1-loop ($\alpha_s^2$) correction 
to the spectator-scattering kernel $T_i^{\rm II}$  
for what is known as the (topological) ``tree amplitudes'' in 
two-body decays. There are several motivations for performing this 
calculation:
\begin{itemize}
\item As in any perturbative QCD calculation the 1-loop correction 
is necessary to eliminate scale ambiguities. In the present case 
of spectator scattering the characteristic scales are $m_b$ and 
$(m_b\Lambda_{\rm QCD})^{1/2}$. The latter being only about
$1.5\,\mbox{GeV}$, a 1-loop calculation is necessary to ascertain 
the validity of a perturbative treatment by showing that the 
expansion converges. The 1-loop correction to spectator scattering 
forms part of the next-to-next-to-leading order (NNLO) contribution to 
the decay amplitudes. 
\item At order $\alpha_s$ the strong interaction phases, and hence 
direct CP asymmetries, originate entirely from the imaginary part of 
the kernel  $T_i^{\rm I}$ in the first term on the right-hand side 
of (\ref{eq:qcdfact}). The 1-loop correction to $T_i^{\rm II}$ 
introduces a new rescattering mechanism by spectator scattering. 
Its calculation represents an important, presumably dominant, part of 
the next-to-leading order (NLO) result for the CP asymmetries. The 
NLO result will be needed to resolve or understand potential 
discrepancies of the LO result with experimental data.
\end{itemize}  

The organization of the paper is as follows. In
Section~\ref{sec:setup} we set up the definitions and 
matching equations for the calculation of the hard-scattering 
kernel $T_i^{\rm II}$, which is then described in 
Section~\ref{sec:calc}. The expression for the kernel is given 
at the end of that section. In Section~\ref{sec:treeamps} we obtain 
the tree amplitudes $\alpha_{1,2}(M_1 M_2)$ 
in a convenient representation, where the light-cone distribution 
amplitudes are integrated in the Gegenbauer expansion. 
The numerical effect of the new correction on the tree 
amplitudes and the $B\to \pi\pi$ branching fractions is 
investigated in Section~\ref{sec:pipi}. We conclude in 
Section~\ref{sec:conclude}.

\section{Set-up and matching}
\label{sec:setup}

\subsection{Flavour and colour}
\label{flc}

\begin{figure}[t]
    \vspace{0.3cm}
\centerline{\includegraphics[width=8cm]{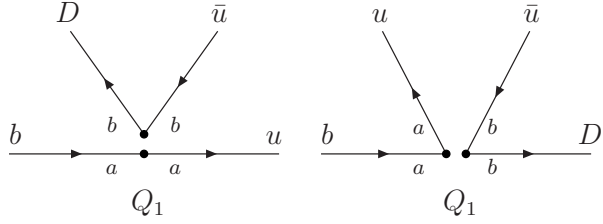}}
\caption{\label{fig1} Flavour and colour flow for insertions 
of $Q_1$. The connected fermion lines indicate the contraction 
of spinor indices. Left: ``right insertion''. Right: ``wrong 
insertion''.}
\end{figure}
We are concerned with the current-current operators in 
the effective weak Hamiltonian for $b\to u$ transitions 
given by 
\bea             
\cH_\eff &=& \frac{G_F}{\sqrt{2}} \,V_{uD}^* V_{ub}
           \left( C_1 Q_1 + C_2 Q_2\right) + \mathrm{h.c.},
\nonumber\\
        &&Q_1 = [\bar u_a \gamma^\mu (1 -\gamma_5) b_a]
                [\bar D_b \gamma_\mu (1 - \gamma_5) u_b] ,
\nonumber\\
        && Q_2 = [\bar u_b \gamma^\mu (1 -\gamma_5) b_a]
                 [\bar D_a \gamma_\mu (1 - \gamma_5) u_b],
\label{eq:weakham}
\eea
with $a,b$ denoting color, and $D=d$ or $s$. There are two possible 
flavour flows to the final state as illustrated in 
Figure~\ref{fig1} for $Q_1$. In case of the colour-allowed tree 
amplitude (left), denoted $\alpha_1(M_1 M_2)$ following the 
notation\footnote{The normalization   
is such that at tree-level $\alpha_1(M_1 M_2)=C_1+C_2/3$ and 
$\alpha_2(M_1 M_2)=C_2+C_1/3$ with $C_1 \sim 1.1$ and 
$C_2\sim -0.2$. See \cite{Beneke:2003zv}, section
2.2, for the relation between decay amplitudes and the $\alpha_i$ 
parameters.} of~\cite{Beneke:2003zv}, meson $M_2$ represented by the up-going 
quark lines has the flavour quantum numbers of $[\bar uD]$, 
and $M_1$ those of $[\bar q_s u]$, where $\bar q_s$ denotes 
the flavour of the spectator anti-quark in the $\bar B$ meson. 
In case of the colour-suppressed tree amplitude $\alpha_2(M_1 M_2)$ (right in
Figure~\ref{fig1}), the corresponding quantum numbers are 
$[\bar u u]$ and $[\bar q_s D]$, respectively. In addition there 
exist ``penguin contractions'', where the $u$ and $\bar u$ 
fields from $Q_i$ are contracted in the same fermion loop. 
Together with other operators 
from the effective Hamiltonian they contribute to the (topological) 
penguin amplitudes, which we do not consider in this paper. 
Thus, in the computation of the (topological) tree amplitudes 
there appear four short-distance coefficients, two corresponding 
to the matrix element of $Q_1$ as shown in Figure~\ref{fig1}, 
and two corresponding to $Q_2$, which differ only by the colour 
labels at the operator vertex. It will be seen from the 
final result that only two of the four coefficients are different, 
because $Q_1$ and $Q_2$ are equivalent by a Fierz transformation, 
when the flavours $u$ and $D$ are not distinguished.  
However, since we use dimensional regularization, Fierz symmetry
cannot be assumed to hold {\em a priori}.

We shall refer to 
the flavour-flow diagrams, where the spinor indices are contracted 
along the quark lines of $M_2$ (left diagram of Figure~\ref{fig1}),  
as the ``right insertions'' of $Q_{1,2}$; the other contraction 
(right diagram) is the ``wrong insertion''. Exactly the same diagrams 
contribute to the two right (wrong) insertions, only the colour 
factor is different for each diagram, since the two operators 
$Q_{1,2}$ have different colour-orderings.  With colour and flavour 
thus understood, we will omit colour and flavour labels in the 
subsequent discussion of operator matching.  

\subsection{Matching onto SCET${}_{\rm I}$}  
\label{sec:ffoview}

The short-distance kernels $T^{\rm I,II}_i$ can be determined 
by extracting the hard and hard-collinear momentum regions from 
quark decay amplitudes according to the strategy of expanding 
Feynman diagrams by regions~\cite{Beneke:1997zp}. The calculation 
becomes more transparent, when it is organized as an operator matching
calculation in soft-collinear effective theory (SCET)~\cite{scet}. 
The spectator-scattering kernel is obtained by the matching
sequence QCD $\to$ SCET${}_{\rm I}$ $\to$ SCET${}_{\rm II}$, 
by which hard fluctuations ($k\sim m_b$, virtuality $m_b^2$) 
and hard-collinear 
fluctuations ($n_+ k\sim m_b$, $k_\perp\sim (m_b\Lambda_{\rm
  QCD})^{1/2}$, $n_- k\sim \Lambda_{\rm QCD}$, virtuality
$m_b\Lambda_{\rm QCD}$) are integrated 
out in two steps. This method has by now been worked out completely 
for heavy-to-light form factors at large recoil energy of the light
meson, both to all orders~\cite{BF03,Bauer:2002aj}, 
and by explicit 1-loop calculations of the short-distance coefficients
\cite{Beneke:2004rc,Hill:2004if,Beneke:2005gs}. For application to 
non-leptonic decays the effective theory has to be extended to 
include two sets of collinear fields corresponding to the 
(nearly) light-like directions of the two final-state mesons. 
As explained in \cite{Chay:2003ju} 
this is a relatively minor complication, because the collinear fields 
for different directions decouple already at the scale $m_b$. 

Our SCET conventions follow those of the 
form-factor calculations \cite{BF03,Beneke:2004rc,Beneke:2005gs}. 
Meson $M_1$, which picks up the spectator anti-quark from the $\bar B$
meson, moves into the direction of the light-like vector $n_-$.  
The collinear quark field for this direction is denoted by 
$\xi$ with $\slash{n}_-\xi=0$, the corresponding 
collinear gluon field is $A_{c1}$. 
The second meson moves into the opposite direction $n_+$, and 
the collinear fields for this direction are $\chi$, satisfying 
$\slash{n}_+\chi=0$, and $A_{c2}$. The heavy quark field $h_v$ 
is labeled by the time-like vector $v=(n_-+n_+)/2$ with $v^2=1$. 

In \cite{BF03} a power-counting argument has been developed to 
identify the SCET${}_{\rm I}$ operators that can appear at leading 
power in the $1/m_b$ expansion of heavy-to-light form factors. 
Applying this argument to the two collinear directions separately, 
we find that $Q_{1,2}$ can match to only two operators in 
SCET${}_{\rm I}$ with non-vanishing matrix elements 
$\langle M_1 M_2| \ldots|\bar B\rangle$ for non-singlet 
mesons $M_{1,2}$.\footnote{Recall that we are not 
counting flavour degrees of freedom. Mesons with 
flavour-singlet components require additional two-gluon operators, as well 
as a term that does not factorize in SCET \cite{Beneke:2002jn}.}
The leading operator in the collinear-2 sector is uniquely 
given by  $(\bar\chi \Wcp)(t n_-) \slash{n}_- (1 - \gamma_5) 
(\Wcp^\dag \chi )(0)$. The two operators are then constructed 
by multiplying this with an A0- and a B1-type current for 
the $\bar B\to M_1$ transition \cite{BF03}. Due to chirality 
conservation and the requirement that the operator be a Lorentz
scalar, there is only one current of each type.  The two SCET${}_{\rm I}$ 
operators thus obtained can be arranged to reproduce the 
structure of the factorization 
formula~\eq{eq:qcdfact} by defining 
\bea
O^{\rm I}(t) &=& 
  (\bar \chi\Wcp) (t n_-) \nmh (1 - \gamma_5) (\Wcp^\dag \chi) \;
        \Big[ \tilde C^{(A0)}_{f_+} \,(\xib W_{c1}) 
        \npslash (1 - \gamma_5) h_v 
\nonumber \\*
&&\hspace*{1cm} 
  - \frac{1}{m_b} \int d\hat{s}\, \tilde C^{(B1)}_{f_+}(\hat{s})\,
    (\bar \xi W_{c1}) \npslash [W^\dagger_{c1}i 
        \Slash{D}_{\perp c1} W_{c1}](s n_+)
                (1 + \gamma_5) h_v \Big] ,
\nonumber \\
\OII(t,s) &=& \frac{1}{m_b} \Big[
   (\bar \chi\Wcp)(t n_-) \nmh (1 - \gamma_5) (\Wcp^\dag\chi)\Big] 
\nonumber \\ 
&&\hspace*{1cm} 
\times \Big[(\bar \xi W_{c1}) \nph [W^\dagger_{c1}i \Slash{D}_{\perp c1}
                 W_{c1}](s n_+) (1 + \gamma_5) h_v\Big].\quad
\label{ops}
\eea
The first operator includes the short-distance coefficients 
$\tilde C^{(A0)}_{f_+}$, $\tilde C^{(B1)}_{f_+}(\hat{s})$ such that 
its matrix element is proportional to the form factor 
$f_+^{BM_1}(0)$ ($A_0^{B M_1}(0)$ for vector mesons) 
in QCD (not SCET${}_{\rm I}$). The expressions 
for the coefficients to 1-loop (more precisely, their momentum space 
Fourier transforms) can be found in \cite{Beneke:2005gs}, 
but they will not be needed here. In (\ref{ops}) fields without 
position argument are at $x=0$, and the field products within the 
large brackets are colour-singlets. We do not consider colour-octet
operators, since their matrix elements between meson states vanish. 
Although the second operator carries 
an apparent $1/m_b$ suppression, both operators are in fact leading, 
because the matrix element of $O^{\rm I}(t)$ is suppressed. 
Hence, at leading order in $1/m_b$, the 
operators $Q_{1,2}$ from (\ref{eq:weakham}) are 
represented in SCET${}_{\rm I}$ by the equation
\begin{equation}
Q = \int d\hat{t}\,\tilde{T}^{\rm I}(\hat t) O^{\rm I}(t) + 
\int d\hat{t}d\hat{s}\,\tilde{H}^{\rm II}(\hat t,\hat s)\OII(t,s)
\label{match1}
\end{equation}
with $\hat s=n_+ p^\prime s= m_B s$, $\hat t=n_- q \,t=m_B t$, 
and $p^\prime$ ($q$) the momentum of $M_1$ ($M_2$). Of the two 
matching coefficients $T^{\rm I}(u) = \int d \hat t \,e^{i u \hat t} 
\,\tilde T^{\rm I}(\hat t)$ is already known to the 1-loop 
order ($\alpha_s$)~\cite{Beneke:1999br}. In this paper we compute 
the 1-loop ($\alpha_s^2$) correction to 
\begin{equation}
H^{\rm II}(u,v) = \int d\hat{t}d\hat{s}\,
e^{i (u\hat{t}+(1-v)\hat {s})}\,
\tilde{H}^{\rm II}(\hat t,\hat s).
\end{equation}
We recall that on accounting for flavour there are actually two 
copies of $\OI(t,s)$, $\OII(t,s)$ with different flavour structure, 
and given the two operators 
$Q_{1,2}$ in the effective Hamiltonian, there are four 
different coefficient functions $H^{\rm II}(u,v)$, which we do 
not distinguish here to simplify the notation.

To see how (\ref{eq:qcdfact}) follows and to make the overall factors 
explicit, we evaluate the matrix element of (\ref{match1}) for the 
case that $M_1$ and $M_2$ are both pseudoscalar mesons. The SCET 
Lagrangian contains no leading-power interactions between 
the collinear-2 and collinear-1 fields after decoupling soft 
gluons from the collinear-2 sector by a field redefinition (second 
paper of \cite{scet}). The matrix elements of 
$\OI(t,s)$, $\OII(t,s)$ fall apart 
into (\cite{Beneke:2005gs}, eqs.~(18,81) with $E=n_+ p^\prime/2=m_B/2$)
\begin{eqnarray}
&& \langle M_2|(\bar \chi\Wcp)(t n_-) \nmh (1 - \gamma_5) (\Wcp^\dag\chi)
  |0\rangle = \frac{i f_{M_2}m_B}{2} 
  \int_0^1 du\,e^{i u \hat{t}} \,\phi_{M_2}(u),
\nonumber \\
&& \langle M_1|(\bar \xi W_{c1}) \nph 
  [W^\dagger_{c1}i \Slash{D}_{\perp c}W_{c1}](s n_+)
  (1 + \gamma_5) h_v|\bar B\rangle = -m_b m_B \int_0^1 d\tau 
  \,e^{i\tau\hat{s}}\,\Xi_{M_1}(\tau),
\label{me1}
\end{eqnarray}
such that
\begin{eqnarray}
\langle M_1 M_2|Q|\bar B\rangle &=& 
  i m_B^2\,\Bigg\{\,f_+^{B M_1}(0) \int_0^1 du\, T^{\rm I}(u)\, 
  f_{M_2} \phi_{M_2}(u) 
\nonumber\\
  &&\hspace*{1.1cm}
   -\,\frac{1}{2} \int_0^1 du dz\,H^{\rm II}(u,z)\,\Xi_{M_1}(1-z)\,
   f_{M_2} \phi_{M_2}(u)\Bigg\}
\label{scet1fact}
\end{eqnarray}
A demonstration of factorization should provide an argument 
for the convergence of the various convolution integrals, an issue 
that is not solved to all orders in perturbation theory for 
the second term (spectator scattering) in the bracket. The convergence 
will be explicitly checked at 1-loop in our calculation. At the 1-loop order 
it is also easy to see by diagrammatic analysis that no operators 
other than $O^{\rm I}(t)$, $O^{\rm II}(t,s)$ are needed to reproduce the 
hard momentum regions. In particular any diagrams that match directly 
onto six-quark operators already in \sceti\ are power-suppressed. 

\subsection{Matching onto SCET${}_{\rm II}$}  

To complete the derivation of (\ref{eq:qcdfact}) the hard-collinear 
scale is integrated out by matching onto SCET${}_{\rm II}$. Hard-collinear 
momentum regions appear only in spectator scattering, since an external 
soft momentum is required. Thus the first term in the bracket of 
(\ref{scet1fact}) is left unchanged, while the SCET${}_{\rm I}$ 
form factor $\Xi_{M_1}$ related to the matrix element (\ref{me1}) 
must be matched onto SCET${}_{\rm II}$. No new calculation is needed 
for this step, since we can use (\cite{Beneke:2005gs}, 
eq.~(86))
\begin{equation}
\Xi_{M_1}(\tau) = \frac{m_B}{4 m_b} \int_0^\infty d\omega 
\int_0^1 dv\,J_\parallel(\tau;v,\omega)\,\hat{f}_B\phi_{B+}(\omega)\,
f_{M_1}\phi_{M_1}(v),
\end{equation}
where the ``jet function'' $J_\parallel(\tau;v,\omega)$ 
has been calculated to 1-loop in 
\cite{Hill:2004if,Beneke:2005gs}, and $\hat{f}_B$ is $\sqrt{m_B}$ 
times the $B$ decay constant in the static limit (\cite{Beneke:2005gs}, 
eq.~(83)). Inserting this into (\ref{scet1fact}), we obtain 
\begin{eqnarray}
\langle M_1 M_2|Q|\bar B\rangle &=& 
  i m_B^2\,\Bigg\{\,f_+^{B M_1}(0) \int_0^1 du\, T^{\rm I}(u)\, 
  f_{M_2} \phi_{M_2}(u) 
\nonumber\\
  &&\hspace*{-1cm}
   + \int_0^\infty d\omega \int_0^1 du dv\,T^{\rm II}(\omega,u,v)\,
     \hat{f}_B\phi_{B+}(\omega)\,
     f_{M_1}\phi_{M_1}(v)\,f_{M_2} \phi_{M_2}(u)\Bigg\},
\label{scet2fact}
\end{eqnarray}
which (up to a normalization factor $i m_B^2$) is (\ref{eq:qcdfact}) with 
\begin{equation}
T^{\rm II}(\omega,u,v) = -\frac{m_B}{8 m_b}
\int_0^1 dz \,H^{\rm II}(u,z)\,J_\parallel(1-z;v,\omega).
\label{t2}
\end{equation}
The jet function is unique, i.e. all four hard-scattering 
functions $H^{\rm II}(u,z)$ are convoluted with the same 
$J_\parallel(1-z;v,\omega)$. 

\begin{figure}[t]
   \vspace{-6cm}
\centerline{\includegraphics[width=26cm]{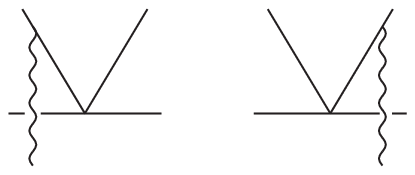}}
   \vspace*{-28cm}
\caption{\label{fig2} Tree diagrams for $H^{\rm II (0)}(u,z)$.}
\end{figure}
The tree-level expressions for the hard coefficient 
functions (when not zero) and the jet function are 
\begin{eqnarray}
H^{\rm II (0)}(u,z) &=& \frac{1}{N_c}\,\frac{2}{\bar u},
\nonumber\\
J_\parallel^{(0)}(\bar z;v,\omega) &=& -\frac{4\pi\alpha_s C_F}{N_c}\,
\frac{1}{m_B\omega\bar v}\,\delta(\bar z-\bar v),
\label{tree1}
\end{eqnarray}
where we introduced the QCD colour factors $C_F=(N_c^2-1)/(2 N_c)$, 
$C_A=N_c=3$, and the ``bar notation'', in which $\bar x\equiv 1-x$ for 
convolution variables $x$. Only the two diagrams shown in 
Figure~\ref{fig2} have to be computed to obtain $H^{\rm II
  (0)}(u,z)$. The other two diagrams with attachments to the horizontal 
quark lines are included in the tree contribution to $T^{\rm I} 
C_{f_+}^{(B1)}$, and thus belong to the $\tilde T^{\rm I} O^{\rm I}$ term 
in (\ref{match1}). Combining the tree coefficients, we obtain 
\begin{equation}
T^{\rm II(0)}(\omega,u,v) =\frac{\pi\alpha_s C_F}{N_c^2}\,
\frac{1}{m_b\omega\bar u \bar v},
\end{equation}
which reproduces the result from~\cite{Beneke:1999br}. Note that  
$m_b$ denotes the $b$-quark pole mass, and $m_B$ the $B$ meson mass, 
but that factors 
of $m_b$ and $m_B$ have not been distinguished in~\cite{Beneke:1999br}, 
since the difference is a power correction. 


\section{1-loop calculation}
\label{sec:calc}

In this section we describe technical aspects of the 
computation of $H^{\rm II}(u,v)$. We calculate the 5-point 
$b\to q_{c2}\bar q_{c2} q _{c1} g_{c1}$ amplitude
\begin{equation}
\langle q(q_1)\bar q(q_2) q(p^\prime_1) g(p_2^\prime)|Q|
b(p)\rangle
\label{amp1}
\end{equation} 
and the corresponding SCET${}_{\rm I}$ matrix elements of the
right-hand side of (\ref{match1}). With the exception of one class of
diagrams to be discussed below, the parton momenta can be restricted 
to their leading components. Thus for the partons in the 
collinear-2 direction we put $q_1=u m_b n_+ /2$, $q_2=\bar u m_b n_+/2$, 
for those in the collinear-1 direction $p_1^\prime =v m_b n_-/2$, 
$p_2^\prime =\bar vm_b n_-/2$, and for the heavy quark momentum 
$p^\mu=m_b v^\mu$. For such external momenta the SCET and 
HQET spinors coincide with the QCD ones.

We use dimensional regularization with $d=4-2\epsilon$ and 
an anti-commuting $\gamma_5$ (NDR scheme). The amplitude (\ref{amp1}) has 
ultraviolet (UV) and infrared (IR) singularities. The former 
must be subtracted in accordance with the definition of the 
operators $Q$ in the effective Hamiltonian \cite{Buras:1989xd}; 
the latter in accordance with the definition of the 
jet function $J_\parallel$ and light-cone distribution 
amplitudes. This is accomplished by using 
$\overline{\rm MS}$ subtractions and a certain 
prescription for dealing with evanescent operators. 
We first discuss the 
``right insertion'' of $Q$, in which the quark spinor indices 
are contracted according to $[\bar \chi\chi] [\bar\xi h_v]$. 
The ``wrong insertion'' leads to  $[\bar\xi \chi] [\bar \chi h_v]$, 
which differs from the desired order (\ref{ops}) by a Fierz 
transformation. 

\subsection{Evanescent operators}

The calculation in dimensional regularization is complicated by the 
presence of evanescent products of Dirac matrices (products 
that vanish in four dimensions). When such products multiply 
$1/\epsilon$ poles they need special treatment. In our calculation 
there are evanescent products that multiply UV singularities. 
Their definition is related to the renormalization convention 
for $Q$. The NDR-$\overline{\rm MS}$ scheme corresponds to 
setting 
\begin{equation}
\gamma^\mu \gamma^\nu \gamma^\rho (1-\gamma_5) 
\otimes \gamma_\mu \gamma_\nu \gamma_\rho  (1-\gamma_5) 
= (16 - 4 \epsilon)\,
  \gamma^\mu  (1-\gamma_5) \otimes \gamma_\mu  (1-\gamma_5), 
\label{uvstring}
\ee
whenever the left-hand side multiplies an UV pole. All other 
products multiplying UV singularities can be reduced to 
(\ref{uvstring}) by permutations.

The evanescent 
products that multiply IR poles are more complicated. Their
treatment is related to the definition of evanescent operators 
of the $O^{\rm II}(t,s)$ type in SCET${}_{\rm I}$. To reduce 
the notation to the essentials, we strip off all the fields 
from  $O^{\rm II}(t,s)$ and represent it only by its Dirac 
structure, 
\begin{equation}
O^{\rm II}(t,s) \to 
\nmh (1 - \gamma_5) \,\otimes\, \nph(1-\gamma_5) \gp^\mu, 
\end{equation}
where 
\begin{equation}
\gamma^{\mu} = \nph n^{\mu}_{-} + \nmh n^{\mu}_{+} +
\gamma^{\mu}_{\perp}.
\end{equation}
In our 1-loop calculation we encounter the four operators 
\bea
  O_{1} &=& \nmh (1 - \gamma_5) \,\otimes\, \nph(1-\gamma_5) \gp^\mu ,
\nonumber\\
  O_{2} &=& \nmh \gp^\mu \gp^\alpha (1-\gamma_5)
            \,\otimes\, \nph(1-\gamma_5)\gamma_{\perp\alpha} ,
\nonumber\\
  O_{3} &=& \nmh \gp^\alpha \gp^\beta (1-\gamma_5) \,\otimes\,
            \nph(1-\gamma_5)  \gp^\mu \gamma_{\perp\alpha} 
            \gamma_{\perp\beta} ,
\nonumber \\
  O_{4} &=&
          \nmh \gp^\mu \gp^\alpha \gp^\beta \gp^\gamma (1-\gamma_5)
          \,\otimes\, \nph(1-\gamma_5) \gamma_{\perp\alpha}
                        \gamma_{\perp\beta} \gamma_{\perp\gamma}.
\label{rightbasis}
\eea
In this notation $O_1$ equals $O^{\rm II}(t,s)$. One easily 
checks that the other three operators are evanescent, i.e. 
vanish in four dimensions. These operators will disappear from 
the final result, since we shall renormalize them such that 
their IR-finite matrix elements vanish, but they must be 
kept in intermediate steps, hence the matching equation 
(\ref{match1}) has to be extended to include all four operators 
on the right-hand side. 

Evanescent operators appear already at tree level. In this
approximation the matrix element (\ref{amp1}) is given by 
\begin{equation}
\langle Q_2\rangle_{\rm nf}  = \frac{1}{N_c}\left(\frac{2}{\bar u}
\langle O_1\rangle - \frac{1}{u\bar u}\langle O_2\rangle
\right).
\label{treeQ}
\end{equation}
(The ``right insertion'' of $Q_1$ vanishes at tree level, because 
the colour-trace is zero.)
The subscript ``nf'' (for ``non-factorizable'') means 
that the ``factorizable'' terms that belong to 
$T^{\rm I} O^{\rm I}$ are omitted, and only the two diagrams 
in Figure~\ref{fig2} are included. 
While one can simply set $\langle O_2\rangle=0$ here to recover 
(\ref{tree1}), since no $1/\epsilon$ poles are present at tree level, the 
appearance of an evanescent operator at tree level implies 
that one must compute the mixing of $O_2$ into $O_1$ in the 
1-loop calculation. 

\subsection{UV renormalized 1-loop amplitude}

\begin{figure}[p]
    \vspace{-6cm}
\centerline{\includegraphics[width=23cm]{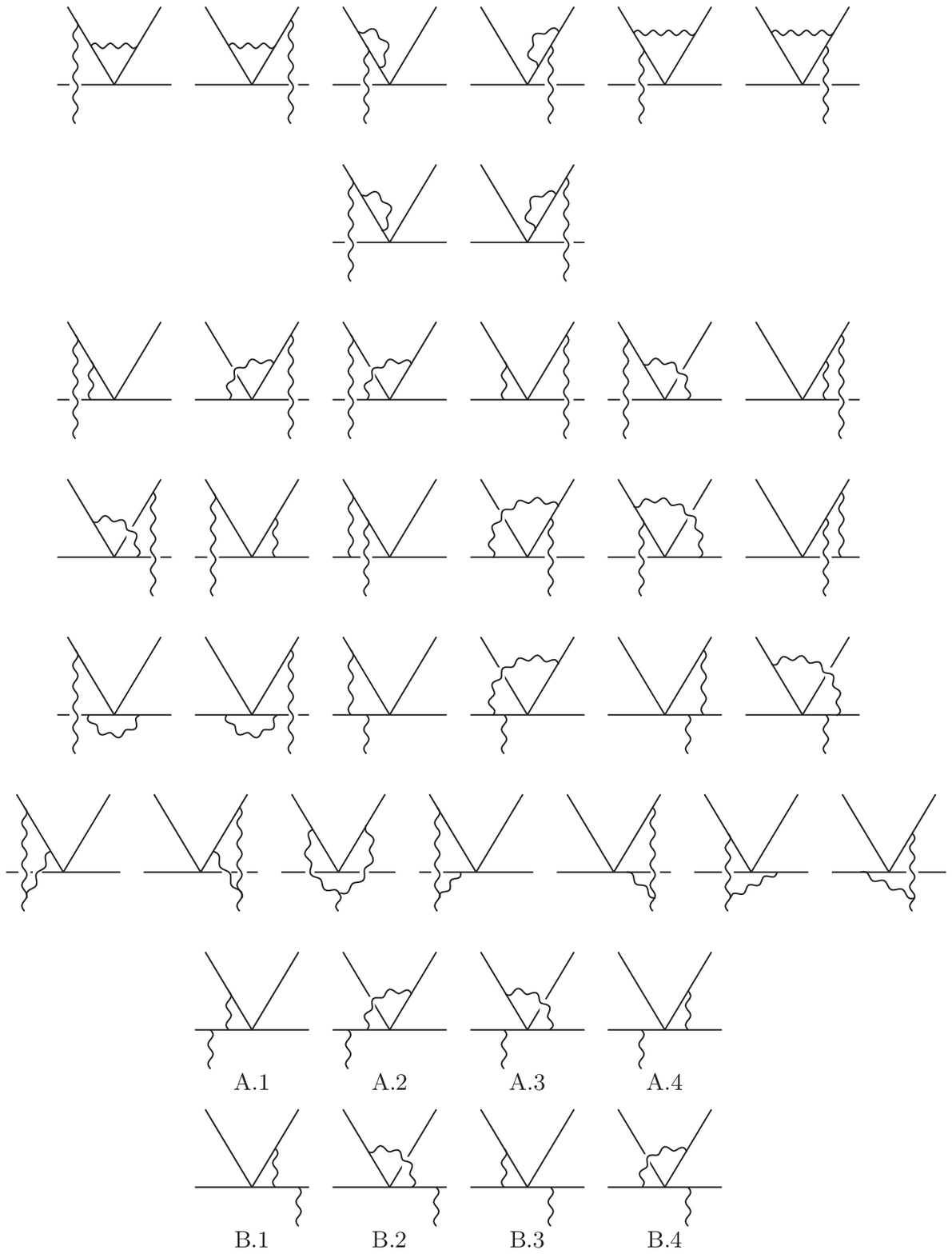}}
    \vspace*{-7cm}
\caption{\label{fig3} 1-loop diagrams for $\langle Q\rangle_{\rm nf}$.}
\end{figure}
The calculation of the 1-loop correction to $H^{\rm II}(u,v)$ 
involves the diagrams shown in Figure~\ref{fig3}.  The two lines 
directed upward represent the quark (anti-quark) with collinear-2 
momentum proportional to $n_+$. The horizontal lines describe an 
incoming bottom quark, and an outgoing collinear-1 quark with momentum 
proportional to $n_-$. The momentum of the external gluon is also 
in the $n_-$ direction. The calculation 
of diagrams with no gluon lines that connect the two upper lines to 
the horizontal lines is 
not necessary, since the definition of $O^{\rm I}$ is chosen such 
that these diagrams contribute only to $T^{\rm I}(u)$.

The calculation of the diagrams uses standard methods. The massive 
box integrals in dimensional regularization can be evaluated adapting 
the method of \cite{Duplancic:2000sk}. Alternatively, they can be reduced to 
vertex integrals, because all external momenta are linear combinations
of only two vectors $n_-$, $n_+$. This observation also simplifies 
the tensor reduction, since one can use 
\be             
        k^\alpha_\perp k^\beta_\perp \to
                \frac{1}{d-2} \,g_\perp^{\alpha\beta} k_\perp^2.
\ee
The classes A, B of 1-particle reducible diagrams must be included 
in the amplitude calculation. The heavy-quark propagator to the right
of the external gluon line in class A is off-shell by an 
amount of order $m_b^2$, hence these diagrams contribute entirely to the 
short-distance coefficient. Class B is more complicated, since 
the light-quark propagator with momentum
$p^\prime=p^\prime_1+p^\prime_2$ has small virtuality, hence 
the diagram is not completely short-distance. The non-local,  
long-distance contributions cancel in the matching relation 
against time-ordered products of $O^{\rm I}$ and the 
SCET interaction Lagrangian as discussed in \cite{Beneke:2004rc}. 
The local contribution to the short-distance coefficient 
can be extracted via the substitution 
\be
\frac{i \slash p'}{{p^\prime}^2} \to \frac{i}{n_+ p^\prime}\,\nph.
\label{sub1}
\ee
A short-cut to this conclusion is obtained by observing that 
we can put $p_{1\perp}^\prime=p_{2\perp}^\prime=p_{\perp}^\prime=0$, 
since we do not match operators with transverse derivatives, and 
keep only $p^{\prime\mu} = n_+ p^\prime \,n_-^\mu/2+
n_- p^\prime\, n_+^\mu/2$. For $p_\perp^\prime=0$ all relevant 
interaction terms from the SCET Lagrangian vanish, hence the class B 
diagrams are purely short-distance. Indeed, since the $\slash{n}_-$ 
term in the propagator does not contribute owing to  the on-shell 
spinor to the right, the substitution (\ref{sub1}) becomes an identity for 
$p_\perp^\prime=0$.

Ultraviolet renormalization of the amplitude involves standard
counterterms from the QCD Lagrangian as well as the counterterms 
for $Q$. The UV-renormalized amplitude is written as 
\be
\langle Q\rangle_{\rm nf} = \sum_{i=1}^4 \left(A_i^{(0)}+A_i^{(1)}\right)
\langle O_i\rangle ^{(0)},
\label{aexp}
\ee
where $\langle O_i\rangle ^{(0)}$ denotes the partonic 
tree-level matrix element of $O_i$, equal to the Dirac matrix 
products (\ref{rightbasis}) multiplied by the SCET quark spinors 
and gluon polarization vector. As already mentioned the tree 
matrix element of the ``right insertion'' of $Q_1$ vanishes 
due to colour, so $A_i^{(0)}=0$ for $Q_1$. In this case the 
1-loop amplitudes  $A_i^{(1)}$ are IR-finite and can be 
evaluated in $d=4$ (after UV renormalization is applied). 
Hence the evanescent terms $i=2,3,4$ vanish. The ``right 
insertion'' of $Q_2$ has $A_i^{(0)} \not=0$ for $i=1,2$ 
[see (\ref{treeQ})], and the 1-loop amplitudes are IR-divergent. 
The $i=1,2$ 1-loop terms have a $1/\epsilon^2$ singularity, proportional 
to the tree matrix element, as follows from the universality of 
soft singularities.  $A_3^{(1)}$ has a $1/\epsilon$ pole, 
while  $A_4^{(1)}$ turns out to be IR-finite. The IR divergences
cancel when the QCD amplitude is related to the matching coefficient 
$H^{\rm II}$ through (\ref{match1}) as explained in the following.

\subsection{IR subtractions}

We start from the matching equation (\ref{match1}) extended 
to include the evanescent operators 
\be 
Q = T^{\rm I}\ast O^{\rm I} +
\sum_i H^{\rm II}_i\ast O^{\rm II}_i.
\label{qt}
\ee
Convolutions, which may involve one or two integrations, 
are now represented by an asterisk. Since we work with matrix elements
in states with definite momentum it is convenient to use the 
momentum-space representation.
Expanding all quantities to the 1-loop order, making use of
(\ref{aexp}) and $O_i^{\rm II} = O_i$, we obtain 
\begin{eqnarray}
\sum_i \left(A_i^{(1)}+A_{i,\rm f}^{(1)}\right)
\langle O_i\rangle ^{(0)} 
&=& T^{{\rm I}(1)}  \langle O^{\rm I} \rangle ^{(0)} 
+ T^{{\rm I}(0)}  \ast\langle O^{\rm I} \rangle ^{(1)} 
\nonumber\\ 
&&\hspace*{0.0cm} +\, 
\sum_i \bigg(H^{{\rm II}(1)}_i  \langle O_i \rangle ^{(0)} 
+ H^{{\rm II}(0)}_i  \ast\langle O_i \rangle ^{(1)} 
\bigg)
\label{match2}
\end{eqnarray}
The factorizable contribution $A_{i,\rm f}^{(1)}$ 
on the left-hand side comes from 1-loop diagrams with no gluon 
lines connecting the $\bar\chi\chi$ part of the diagram to 
the $\bar\xi \slash{A}_{\perp c1} h_v$ part. It is canceled by the term 
$T^{{\rm I}(0)}  \ast \langle O^{\rm I} \rangle ^{(1)}$ 
on the right-hand side, since the 1-loop matrix element of 
$O_i^{\rm I}$ contains exactly these diagrams in the coefficient 
function $C_{f_+}^{(B1)}$ in its definition (\ref{ops}). The 
UV-renormalized 1-loop matrix elements of the $O_i$ are given by 
\be
\langle O_i \rangle ^{(1)} = 
\sum_j \Big(M_{ij}^{(1)R} + Z_{ij}^{(1)}\Big) \, 
\langle O_i \rangle ^{(0)}, 
\label{1loopme}
\ee
where $M_{ij}^{(1)R}$ is the bare matrix element, which 
depends on the IR regularization scheme $R$, and 
$Z_{ij}^{(1)}$ the matrix kernel of ultraviolet renormalization factors. 
When dimensional regularization is used for UV and IR singularities 
as was done in the calculation of $A_{i}^{(1)}$, the bare 
matrix elements vanish, since the 1-loop diagrams are scaleless. 
Hence, inserting (\ref{1loopme}) into (\ref{match2}), using 
(\ref{ops}) and $H^{{\rm II}(0)}_i = A_i^{(0)}$,  
we obtain 
\be 
A_{1}^{(1)} = H_1^{\rm II (1)} +\sum_{i=1}^2  A_i^{(0)}\ast
Z_{i1}^{(1)} + (-2) T^{\rm I(1)} C_{f_+}^{(B1)(0)}
\label{aexp2}
\ee
by comparing the coefficient of $\langle O_1 \rangle ^{(0)}$. 
We also used that $A_i^{(0)}$ is zero for $i=3,4$. The renormalization
constants for the evanescent operators are determined by requiring 
that the IR-finite matrix elements $\langle O_i \rangle$ 
($i=2,3,4$) vanish \cite{Buras:1989xd, Dugan:1990df}. 
Here ``IR-finite'' means the matrix element 
computed with any IR regularization $R=\,$off  
other than dimensional and with dimensional regularization applied
only to the UV singularities. According to (\ref{1loopme}) 
this fixes $Z_{21}^{(1)} = - M_{21}^{\rm (1) off}$. 
Hence the 1-loop short-distance coefficient of the physical 
(non-evanescent) operator $O^{\rm II}$ is given by 
\be 
H^{\rm II (1)} = A_{1}^{(1)} - A_1^{(0)}\ast Z_{11}^{(1)} + 
A_2^{(0)}\ast M_{21}^{\rm (1) off} + 2 \,T^{\rm I(1)} C_{f_+}^{(B1)(0)}.
\label{fin1}
\ee
Note that since the IR-finite matrix elements of the 
evanescent operators have been made to vanish, only the 
term $i=1$ survives in (\ref{qt}). It is therefore not 
necessary to determine the coefficient functions $H^{\rm II}_i$ 
for $i\not =1$. We also note that the renormalization constant 
$Z_{21}^{(1)}$ is finite, and that $M_{21}^{\rm (1) off}$ 
is independent of the apparently arbitrary IR regulator. This 
is because the mixing of an evanescent operator into a physical 
operator arises through the multiplication of an ultraviolet 
$1/\epsilon$ pole with a term of order $\epsilon$ from the 
Dirac algebra, both of which are independent of the IR
regularization. The $1/\epsilon^2$ poles do not contribute 
to operator mixing due to their universality.

Eq.~(\ref{fin1}) provides the final result for the two of the four 
matching coefficients associated with the right insertions. 
We briefly discuss the subtraction terms in (\ref{fin1}). First 
note that for the right insertion of $Q_1$ the tree amplitudes 
$A_i^{(0)}$ vanish, hence (\ref{fin1}) is simply 
$H^{\rm II (1)} = A_{1}^{(1)}$. This is consistent, since 
for this case $A_{1}^{(1)}$ is IR-finite as observed above. 
For the right insertion of $Q_2$, all three subtraction terms 
in (\ref{fin1}) are present. Due to factorization in 
SCET, the renormalization of an operator $[\bar\chi \chi]\,
[\bar\xi A_{\perp c1} h_v]$ falls apart into a renormalization 
factor for the collinear-2 bracket $[\bar\chi \chi]$ and 
one for a B1-type current. $Z_{11}^{(1)}$ is therefore determined 
by the requirement that the light-cone distribution amplitude 
of $M_2$ and the jet function are defined in the $\overline{\rm MS}$ 
scheme. This gives  $Z_{11}^{(1)}$ as the product of the
Brodsky-Lepage kernel \cite{Lepage:1980fj} and the renormalization 
kernel $Z_\parallel$ for the B1-type current (first paper 
of~\cite{Hill:2004if}, \cite{Beneke:2005gs}). Subtracting 
$A_1^{(0)}\ast Z_{11}^{(1)}$ from $A_{1}^{(1)}$ removes  
the IR singularities such that $H^{\rm II (1)}$ is finite 
as must be for a short-distance coefficient. The third term 
on the right-hand side involves the computation of the 1-loop 
matrix element of the evanescent operator $O_2$. We find that 
only a single diagram, shown in Figure~\ref{fig4}, contributes 
to $M_{21}^{\rm (1) off}$, such that $M_{21}^{\rm (1) off}$ 
is proportional to the spin-dependent part of the 
Brodsky-Lepage kernel. Explicitly, 
\begin{eqnarray}
A_2^{(0)}\ast M_{21}^{\rm (1) off} &=& 
\frac{\alpha_s C_F}{4\pi} \int_0^1 du^\prime \,
\frac{(-1)}{N_c u^\prime \bar u^\prime}\,
(-8) \left(\frac{u^\prime}{u}\,\theta(u-u^\prime)+
\frac{\bar u^\prime}{\bar u}\,\theta(u^\prime-u)\right)
\nonumber\\
&=& \frac{\alpha_s }{4\pi}\frac{C_F}{N_c} \,(-8)
\left(\frac{\ln u}{\bar u}+\frac{\ln \bar u}{u}\right)
\label{sub11}
\end{eqnarray}
The fourth term $2 \,T^{\rm I(1)} C_{f_+}^{(B1)(0)}$ follows 
from $C_{f_+}^{(B1)(0)}=-1$ \cite{Beneke:2005gs} and \cite{Beneke:1999br}
\be
T^{\rm I (1)} = \frac{\alpha_s}{4\pi}\,\frac{C_F}{N_c} \,V(u)
\label{sub12}
\ee
with $V(u)$ given in (\ref{FV}) below. 
Note that $H^{\rm II (1)}(u,v)$ is a function of two variables, 
but like the tree contribution the two subtraction terms 
(\ref{sub11}), (\ref{sub12}) depend on $u$, but not on 
the momentum fraction $v$ related to the collinear-1 momenta. 
\begin{figure}[t]
   \vspace{0cm}
\centerline{\includegraphics[width=3.5cm]{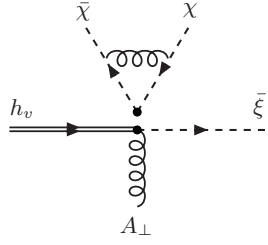}}
   \vspace*{0cm}
\caption{\label{fig4} SCET 1-loop diagram contributing to 
$M_{21}^{\rm (1) off}$, the mixing of $O_2$ into $O_1$.}
\end{figure}

\subsection{``Wrong insertion''}

The other two matching coefficients are related to the wrong 
insertions of $Q_{1,2}$ as in the right diagram of Figure~\ref{fig1}. 
We would like to express them as the coefficients of the same 
SCET${}_{\rm I}$ operator $O^{\rm II}(t,s)$, but the QCD calculation 
involves Dirac matrix products with a different contraction of spinor 
indices corresponding to 
\bea
\tilde O^{\rm II}(t,s) &=& \frac{1}{m_b} \Big[
   (\bar \xi W_{c1})(t n_-) [W^\dagger_{c1}i \Slash{D}_{\perp c1}
                 W_{c1}](s n_+) (1 - \gamma_5) (\Wcp^\dag\chi)\Big] 
\,\Big[(\bar \chi \Wcp) (1 + \gamma_5) h_v\Big]
\nonumber \\ 
&\to & \gp^\mu (1-\gamma_5) \;\sotimes\; (1+\gamma_5)
\label{ops2}
\eea
In the second line we introduced again a short-hand notation 
that highlights the Dirac structure. The symbol $\sotimes$ 
means that the spinor indices are contracted as in $[\bar \xi\chi]\,
[\bar \chi h_v]$. We deal with the required Fierz transformation 
and evanescent operators simultaneously by introducing the operators 
\bea 
  \tilde O_{0} &=& \nmh (1 - \gamma_5) \,\otimes\, \nph(1-\gamma_5) \gp^\mu ,
  \nonumber\\
  \tilde O_{1} &=&\gp^\mu (1-\gamma_5) \;\sotimes\; (1+\gamma_5) ,
  \nonumber\\
  \tilde O_{2} &=& \gp^\alpha (1-\gamma_5) \;\sotimes\;
                (1+\gamma_5)\gamma_\perp^{\mu} \gamma_{\perp\alpha}  ,
  \nonumber\\
  \tilde O_{3} &=&  \gp^\mu \gp^\alpha \gp^\beta (1-\gamma_5) \;\sotimes\;
                 (1+\gamma_5)\gamma_{\perp\alpha} \gamma_{\perp\beta}
                ,
  \nonumber\\
  \tilde O_{4} &=& \gp^\alpha \gp^\beta \gp^\gamma (1-\gamma_5) \;\sotimes\;
       (1+\gamma_5)\gp^{\mu} \gamma_{\perp\alpha} 
       \gamma_{\perp\beta} \gamma_{\perp\gamma}.
\eea
Here $\tilde O_0$ is the short-hand for $O^{\rm II}$. The basis is 
chosen such that $\tilde O_1$ and $\tilde O_0$ are Fierz-equivalent 
and $\tilde O_{2-4}$ vanish in four dimensions. Hence we have one 
physical operator, $\tilde O_0$, and four evanescent operators, 
$\tilde O_1-\tilde O_0$, and $\tilde O_{2-4}$. The tree matrix 
element is now given by 
\begin{equation}
\langle Q_1\rangle_{\rm nf}  = \frac{1}{N_c}\left(\frac{2}{\bar u}
\langle \tilde O_1\rangle +\frac{1}{u}\langle \tilde O_2\rangle
\right),
\label{treeQ2}
\end{equation}
and the 1-loop amplitude reads 
\be
\langle Q\rangle_{\rm nf} = \sum_{i=1}^4 \left(\tilde A_i^{(0)}+
\tilde A_i^{(1)}\right)
\langle \tilde O_i\rangle ^{(0)}.
\label{aexp3}
\ee
This does not contain $\langle \tilde O_0\rangle ^{(0)}$, since all 
diagrams have the ``wrong'' Fierz-ordering. Proceeding as before 
and requiring that the infrared-finite matrix elements of the four 
evanescent operators vanish, we find that (\ref{fin1}) is replaced 
by 
\be 
H^{\rm II (1)} = \tilde A_{1}^{(1)} - 
\tilde A_1^{(0)}\ast \tilde Z_{00}^{(1)} + 
\tilde A_2^{(0)}\ast \tilde M_{21}^{\rm (1) off} + 
\tilde A_1^{(0)}\ast \left(\tilde M_{11}^{\rm (1) off}-
      \tilde M_{00}^{\rm (1) off}\right) + 
2 \,T^{\rm I(1)} C_{f_+}^{(B1)(0)}
\label{fin2}
\ee
with $\tilde M_{21}^{\rm (1) off}$ the bare 1-loop mixing of $\tilde O_2$ 
into $\tilde O_1$, and $\tilde Z_{00}^{(1)} = Z_{11}^{(1)}$. 
The new term $\tilde A_1^{(0)}\ast (\tilde M_{11}^{\rm (1) off}-
      \tilde M_{00}^{\rm (1) off})$ involves the difference 
of the mixing of $\tilde O_1$ and $\tilde O_0$ into themselves. 
This difference is finite and independent of the IR regulator 
for the same reason that $\tilde M_{21}^{\rm (1) off}$ is. 
There is one subtle aspect hidden in (\ref{fin2}) that requires
explanation. As in (\ref{match2}) we would like to cancel the 
factorizable QCD diagrams against the matrix element of $O^{\rm I}$, 
but the two terms appear in different Fierz-orderings. The 
consequence of this is that there should be an extra term related 
to the factorizable diagrams on the right-hand side of (\ref{fin2}). 
Using that at tree-level only $A_{1, \rm f}^{(0)}$, 
$\tilde A_{1, \rm f}^{(0)}$, $\tilde A_{2, \rm f}^{(0)}$ are 
non-zero, it is given by
\be
\tilde A_{1,\rm f}^{(1)} -A_{1,\rm f}^{(1)} + 
\tilde A_{2, \rm f}^{(0)}\ast \tilde M_{21}^{\rm (1) off} + 
\tilde A_{1, \rm f}^{(0)}\ast \left(\tilde M_{11}^{\rm (1) off}-
      \tilde M_{00}^{\rm (1) off}\right).
\ee
However, we find that this term vanishes, hence (\ref{fin2}) is correct. 
The subtractions $-\tilde A_1^{(0)}\ast \tilde Z_{00}^{(1)}$ and 
$2 \,T^{\rm I(1)} C_{f_+}^{(B1)(0)}$ are identical to the
corresponding terms for the right insertion. The other two terms 
are once again related to an integral over the spin-dependent 
part of the Brodsky-Lepage kernel. Explicitly, they read 
\bea
\tilde A_2^{(0)}\ast \tilde M_{21}^{\rm (1) off} 
&=& \frac{\alpha_s }{4\pi}\frac{C_F}{N_c}\,(-8)\,\frac{\ln u}{\bar u},
\nonumber\\
\tilde A_1^{(0)}\ast \left(\tilde M_{11}^{\rm (1) off}-
      \tilde M_{00}^{\rm (1) off}\right)
&=&  \frac{\alpha_s }{4\pi}\frac{C_F}{N_c} \,(-4)\,\frac{\ln \bar
  u}{u}.
\qquad
\eea
Despite the different Dirac algebra and subtraction structure we find 
that the final result for the matching coefficient related to the wrong 
insertion of $Q_{1}$ ($Q_2$) is identical to the one for the right 
insertion of $Q_{2}$ ($Q_1$). 

\subsection{Matching coefficients}

Here we give the final results for the matching coefficients 
(hard spectator-scattering kernels) $H^{\rm II}(u,v)$  
in the convolutions (\ref{scet1fact}), (\ref{t2}). The kernels 
are independent of the mesons but dependent on their flavour 
quantum numbers. To make them explicit, we write 
$\langle [\bar q_s q_{M_1}] [\bar q_{M_2} q_{M_2}]|Q| 
[\bar q_s b]\rangle$ for $\langle M_1 M_2 |Q|\bar B \rangle$ 
to indicate flavour. Then 
\begin{itemize}
\item[--] for $\langle [\bar q_s D] [\bar u u]|Q_1| 
[\bar q_s b]\rangle$, the contribution of $Q_1$ to the
colour-suppressed 
tree amplitude $\alpha_2(M_1 M_2)$, and  
for $\langle [\bar q_s u] [\bar u D]|Q_2| 
[\bar q_s b]\rangle$, the contribution of $Q_2$ to the
colour-allowed tree amplitude $\alpha_1(M_1 M_2)$, we have 
\begin{equation}
H_1^{\rm II}(u,v) = \frac{2}{N_c} \left(\frac{1}{\bar u}+
\frac{\alpha_s}{4\pi}\,r_1(u,v)\right)
\label{ke1}
\end{equation} 
\item[--] for $\langle [\bar q_s u] [\bar u D]|Q_1| 
[\bar q_s b]\rangle$, the contribution of $Q_1$ to the colour-allowed
tree amplitude $\alpha_1(M_1 M_2)$, and for 
$\langle [\bar q_s D] [\bar u u]|Q_2| 
[\bar q_s b]\rangle$, the contribution of $Q_2$ to the
colour-suppressed tree amplitude $\alpha_2(M_1 M_2)$: 
\begin{equation}
H_2^{\rm II}(u,v) = \frac{2}{N_c}\,\frac{\alpha_s}{4\pi}\,r_2(u,v)
\label{ke2}
\end{equation} 
\end{itemize}
Here $r_1(u,v)$, $r_2(u,v)$ are given by 
\begin{eqnarray}
  r_1 &=& C_F\Bigg[
        - \frac{1}{2 \ub} \ln^2 \frac{m_b^2}{\mu^2}
        + \bigg( 6 - \frac{5}{2 \ub} + \frac {2}{\ub} \ln \ub \bigg)
                \ln \frac{m_b^2}{\mu^2}
        + \frac{u}{\ub} \,\Big[ V(u) + 18\Big]
\nonumber \\ && \hspace*{0.8cm}
        - \,\frac{2 u}{\ub} \,F(v, u)
        + \frac{2 u^3}{(\vb-u)^3} \,F(v, \ub)
        - \frac{2}{\ub} \Big( \ln \ub + \ln \vb \Big) i \pi
                 - \frac{1}{\ub} \bigg(9 + \frac{5}{12} \pi^2 \bigg)
\nonumber \\ &&  \hspace*{0.8cm}
        + \,\bigg( \frac{u}{\vb-u} - \frac{2 u^2}{(\vb-u)^2}
        - \frac{2 (3 u - 2)}{\ub} \bigg) \Big[ \log u - i \pi \Big]
        - \frac{2(1- u v)}{u \ub v \vb} \ln (1 - u v)
\nonumber \\ && \hspace*{0.8cm}
        - \,\bigg( \frac{u (2 - 4 u + u^2)}{\ub^2 (\vb-u)}
                - \frac{2 (2-u) u^2}{\ub (\vb-u)^2}
                + \frac{\ub - v + 4 u \ub v}{u \ub^2 v^2} \bigg)
            \ln (1 - \ub v)
\nonumber \\ && \hspace*{0.8cm}
        + \,\frac{1}{\ub} \big( \ln^2 \ub - \ln^2 \vb \big)
        + \bigg( \frac{u}{\vb-u} - \frac{2 u^2}{(\vb-u)^2} +
                \frac{\vb + 3 u v^2}{u v^2} \bigg) \ln \vb
\nonumber \\ && \hspace*{0.8cm}
        + \,\bigg( 3 + \frac{2}{u \vb} - 2 \ln v
                + \Big( 2 + \frac{2}{\ub}\Big) \ln \vb  \bigg) \ln \ub
        + \bigg( 3 + \frac{2}{\ub} \ln \vb \bigg) \ln v
\nonumber \\ && \hspace*{0.8cm}
        + \,\bigg( \frac{1- 3 u \ub}{\ub^2} + \frac{2 (3u-2)}{\ub}
                + 2 \ln v + \frac{2 u}{\ub} \ln \vb \bigg) \ln u
        + \frac{2}{\ub} \,\LL(\vb)\Bigg]
\nonumber \\ &&
+ \left(C_F-\frac{C_A}{2}\right) \Bigg[
        - \frac{2}{\ub\vb}\,\ln v\, \ln \frac{m_b^2}{\mu^2}
        + \frac{1}{\ub}\,V(u)
                + \frac{2u}{v-u} F(v,u) + \frac{2 u^2}{(\vb-u)^2} F(v, \ub)
\nonumber \\ && \hspace*{0.8cm}
        +  \,2 i\pi\,\bigg( \frac{\vb}{\vb-u} + \frac{u}{\ub} \ln \frac{u}{\ub}
                + \frac{v}{\ub \vb} \ln v + \frac{1}{\ub} \ln \vb \bigg)
        + \frac{1}{\ub}  \bigg[ \ln^2 \vb + \frac{\pi^2}{3} \bigg]
        + \frac{u}{\ub} \big( \ln^2 u - \ln^2 \ub \big)
\nonumber \\ && \hspace*{0.8cm}
        - \,\frac{1+\vb}{\ub \vb} \ln^2 v
        +  \bigg( \frac{2-3u}{u\ub} + \frac{2}{u\vb} \bigg) \ln \ub
        - \frac{2(1-u v)}{u\ub v\vb} \ln(1-u v)
\nonumber \\ && \hspace*{0.8cm}
        + \bigg( \frac{2(1-3 u^2 +u^3)}{u \ub v} 
              + \frac{2(1-\vb^2\ub)}{\ub v\vb}
                + \frac{2(2-u)u}{v(\vb-u)} \bigg) \ln (1 - \ub v)
\nonumber \\ && \hspace*{0.8cm}
        + \,\bigg( \frac{1+2u \bar u}{\ub^2} - \frac{2(1-u v)}{\ub^2 \vb}
                + \frac{2u(u \ub - v)}{\ub^2 (\vb-u)} \bigg) \ln u
\nonumber \\ && \hspace*{0.8cm}
        + \,\bigg( \frac{3(1+\vb)}{\ub\vb}
                + \frac{2}{\ub} \ln u - \frac{2}{\ub\vb} \ln \ub \bigg) \ln v
\nonumber \\ && \hspace*{0.8cm}
        - \,\bigg( \frac{2-5u-4u^2+2u^3}{u \ub v} + \frac{(1+2u)\vb}{\ub v}
                + \frac{2 u \ub}{v(\vb-u)} +\frac{2 u}{\ub} \ln u
\nonumber \\ && \qquad \qquad
                + 2 \ln \ub + \frac{2(1-2u)}{u\ub\vb} \ln v \bigg) \ln \vb
\nonumber \\ && \hspace*{0.8cm}
        + \,2\, \bigg\{\! - \frac{1 + 2 u \vb}{\ub \vb} \,\LL(u)
        + \frac{1 - u \vb + 2 u^2 \vb}{u\ub\vb} \,\LL(\ub)
        + \frac{1 + u \vb}{\ub \vb} \,\LL(u v)
\nonumber \\ && \qquad \qquad
        - \,\frac{1 - u \ub \vb}{u \ub \vb} \,\LL(\ub v)
        - \frac{1 - 2 u + u \vb}{u\ub\vb} \,\LL(\vb) \bigg\}\Bigg],
\label{kernel1}
\end{eqnarray}
\begin{eqnarray}
  r_2 &=& \frac{1}{2 \ub} \,\Big[ V(u) + 2 \Big]
    + \frac{u \vb}{\ub (v-u)}  \,F(v, u)
    + \frac{u^2 \vb}{(\vb-u)^3} \,F(v, \ub)
    + \frac{\vb (\vb -3 u)}{2 (\vb-u)^2} \,\Big[ \ln \vb - i \pi \Big]
\nonumber \\ &&
    +\, \bigg(\frac{1 + u^2}{2 \ub^2} + \frac{u}{2 \ub^2\, \vb}
              - \frac{u^2}{(\vb-u)^2} - \frac{u}{2 (\vb - u)} \bigg) \ln u
    + \frac{\ub + \vb}{u \ub \vb} \,\ln \ub
\nonumber \\ &&
    +\, \bigg(\frac{3}{2 \ub} - \frac{1}{2 \ub \vb} + \frac{\ln u}{\ub}
              - \frac{\ln \ub}{\ub} \bigg) \ln v
    - \bigg(\frac{1}{u \ub v} + \frac{1}{u \vb} \bigg) \ln (1 - u v)
\nonumber \\ &&
    +\, \bigg(\! - \frac{1}{2} - \frac{u}{2 \ub^2 \vb}
        + \frac{u^2 (\ub^2 + v)}{\ub^2(\vb-u)^2}
        + \frac{u (2 + u^2)}{2 \ub^2 (\vb-u)} \bigg) \ln (1 - \ub v),
\label{kernel2}
\end{eqnarray}
where we defined  
\begin{eqnarray}
  F(v, w) &=& 2\, \LL\bigg(\!-\frac{\vb w}{\wb}\bigg) + 2\, \LL(w) -
  \LL(v w) + \frac{1}{2} \ln^2 \frac{\wb}{\vb} + 
  i \pi \ln \frac{\bar w}{\bar v} ,
\nonumber \\
  V(u) &=& 6 \ln \frac{m_b^2}{\mu^2} - 18
                + 3 \,\bigg( \frac{1-2 u}{\ub} \ln u - i \pi \bigg)
\nonumber \\ && \quad
        + \bigg\{2\, \LL(u) - \ln^2 u + \frac{2 \ln\,u}{\ub}
                - ( 3 + 2 i \pi) \ln u
                - (u \leftrightarrow \ub) \bigg\} .
\label{FV}
\end{eqnarray}
The expressions for $r_1$ and $r_2$ constitute the main technical 
results of this paper. In applications the kernels always appear in 
convolutions. In the following we perform the convolution integrals 
analytically and obtain a compact representation for the (topological) 
tree amplitudes.

\section{Tree amplitudes with NLO spectator scattering }
\label{sec:treeamps}

The complete 1-loop correction to spectator scattering is the
convolution of the hard-scattering kernels $H^{\rm II}_{1,2}$ with 
the jet function 
\be 
J_\parallel(\bar z;v,\omega) =  -\frac{4\pi\alpha_s C_F}{N_c}\,
\frac{1}{m_B\omega\bar v}\left[\delta(\bar z-\bar v) +  
\frac{\alpha_s}{4\pi}\,j_\parallel(\bar z;v,\omega)\right].
\ee
Inserting  this and (\ref{ke1}), (\ref{ke2}) into (\ref{t2}), 
and expanding to order $\alpha_s^2$, we obtain 
\begin{eqnarray}
T_i^{\rm II}(\omega,u,v) &=& \frac{\pi\alpha_s C_F}{N_c^2}\,
\frac{1}{m_b \omega\bar v}\,
\left\{
\begin{array}{l}
\displaystyle
\frac{1}{\bar u}+
\frac{\alpha_s}{4\pi}\left[r_1(u,v)+\frac{1}{\bar u}
\int_0^1 dz\,j_\parallel(z;v,\omega)\right]\quad (i=1)
\\[0.5cm]
\displaystyle
\frac{\alpha_s}{4\pi}\,r_2(u,v)\quad (i=2)
\end{array}
\right.
\end{eqnarray}
(The integral $\int_0^1 dz\,j_\parallel(z;v,\omega)$ is given 
analytically in appendix B.1 of \cite{Beneke:2005gs}.)
The spectator-scattering contribution to the tree decay amplitudes 
$\alpha_{1,2}(M_1 M_2)$ in the standard normalization is 
\be 
S_i =  \frac{\hat{f}_B f_{M_1}}{f_+^{BM_1}(0)}
\int_0^\infty d\omega \int_0^1 du dv\,T_i^{\rm II}(\omega,u,v)\,
     \phi_{B+}(\omega)\,
     \phi_{M_1}(v)\, \phi_{M_2}(u), 
\label{sdef}
\ee 
see (\ref{scet2fact}). More precisely, accounting for the Wilson 
coefficients in the effective weak Hamiltonian (\ref{eq:weakham}), 
we have $\alpha_1(M_1 M_2)|_{\rm sp} = C_1 S_2+C_2 S_1$, 
$\alpha_2(M_1 M_2)|_{\rm sp} = C_1 S_1+C_2 S_2$. 

\subsection{Expansion of convolutions in Gegenbauer moments}

The integration over the spectator quark momentum fraction 
$\omega$ is simple, because $\omega$ appears only in the jet function,
or as the over-all factor $1/\omega$. The dependence on the 
light-cone distribution amplitude of the $B$ meson is encoded 
in the inverse moment
\begin{equation}
\frac{1}{\lambda_B} \equiv \int_0^\infty
\frac{d\omega}{\omega}\,\phi_{B+}(\omega),
\label{lamb}
\end{equation}
and the logarithmic moments 
\be
  \langle L^n \rangle = \lambda_B \int \frac{d\omega}{\omega}
                        \phi_{B_+}(\omega) \ln^n \frac{m_b \omega}{\mu^2}
\ee
up to $n=2$. 
The light-cone distribution amplitude of a light meson, 
$\phi_M$, is conventionally expanded into the eigenfunctions of 
the 1-loop renormalization kernel, 
\begin{equation}
\phi_M(x)\,=\, 6 x\bar x \left[1+\sum\limits_{n=1}^{\infty}
a_n^M C^{(3/2)}_n(2 x-1)\right]\,,
\end{equation}
where $a_n^M$ and $C^{(3/2)}_n$ are the Gegenbauer
moments and polynomials, respectively. The integrals over 
$u$ and $v$ can then be performed, the result being represented 
as a double expansion in the Gegenbauer moments of $M_1$ and 
$M_2$. Often, there appear the quantities
\begin{equation}
\Delta_M \equiv \int_0^1 dx \,\frac{\phi_M(x)}{3 x} 
= 1 +\sum_{n=1}^\infty (-1)^n\,a_n^M,
\quad 
\bar\Delta_M \equiv \int_0^1 dx \,\frac{\phi_M(x)}{3 \bar x} 
= 1 +\sum_{n=1}^\infty a_n^M,
\end{equation}

We give the final result for the tree amplitude parameters 
$\alpha_{1,2}(M_1 M_2)$ in the notation of \cite{Beneke:2003zv}
[eq.~(35)], including the 1-loop vertex correction $V_i(M_2)$ not 
related to spectator scattering, 
\bea 
\alpha_i(M_1 M_2) &=& C_i+\frac{C_{i\pm 1}}{N_c} + 
\frac{C_{i\pm 1}}{N_c}\,\frac{\alpha_s C_F}{4\pi}\,V_i(M_2) 
\nonumber\\
&&+\,\frac{\pi\alpha_s C_F}{N_c^2}\,\frac{9 f_{M_1}\hat f_B}
{m_b f_+^{BM_1}(0)\lambda_B}\,
\Big[C_{i\pm 1} \,h_1(M_1 M_2) + C_i \,h_2(M_1 M_2)\Big].
\label{alphapar}
\eea
The upper signs apply when $i$ is odd (here simply $i=1$), 
the lower ones when $i$ is even (here $i=2$). The spectator-scattering 
mechanism is encoded in the two objects
\bea
h_1(M_1 M_2) &=& \bar \Delta_{M_1}\bar \Delta_{M_2}+
\frac{1}{3} \,r_\chi^{M_1}\Delta_{M_2} X_H+
\frac{\alpha_s}{4\pi}\,\Big[R_1(M_1 M_2)+\bar \Delta_{M_2} J(M_1)\Big],
\nonumber\\
h_2(M_1 M_2) &=& \frac{\alpha_s}{4\pi} R_2(M_1 M_2),
\label{hs}
\eea
such that the $\alpha_s$ terms in these expressions extend the 
result given in \cite{Beneke:2003zv}, and $r_\chi^{M_1}\Delta_{M_2}
X_H$ denotes a power correction included in the definition 
of $H_i(M_1 M_2)$ in  \cite{Beneke:2003zv}. Performing the 
convolution integrals in a double Gegenbauer expansion as 
described above, the hard-collinear 
1-loop ($\alpha_s^2$) correction is given up to the 
second Gegenbauer moment in terms of (\cite{Beneke:2005gs}, 
appendix B.1) 
\bea
J(M_1)&\equiv& 
\frac{\lambda_B}{3} \int_0^1 \frac{dv}{\bar v}
\,\phi_{M_1}(v) \int_0^\infty \frac{d\omega}{\omega}\,
\phi_{B+}(\omega)\int_0^1 d\tau j_\parallel(\tau;v,\omega)  \nonumber \\
&=& \frac{4}{3} \langle L^2\rangle  + 
\left (-\frac{19}{3}+\frac{\pi^2}{9}\right )
\langle L \rangle
+\frac{169}{18}-\frac{2\pi^2}{9}-\frac{8}{3}\zeta(3)\nonumber\\
&& +a_1^{M_1}\bigg [
\frac{4}{3} \langle L^2\rangle
+\left (-\frac{110}{9}+\frac{\pi^2}{3}\right ) \langle L \rangle
+\frac{464}{27}+\frac{\pi^2}{9}-8\zeta(3)\bigg ]
\nonumber\\
&& +a_2^{M_1}\bigg [ \frac{4}{3} \langle L^2\rangle +\left
(-\frac{157}{9}+\frac{2\pi^2}{3}\right ) \langle L \rangle +
\frac{646}{27}+\frac{8\pi^2}{9}-16\zeta(3)\bigg ]. 
\eea
(Here we have set $n_f=4$, $T_f=1/2$ and $C_F=4/3$ and $C_A=3$.)
The new hard correction is in 
\be
R_k(M_1 M_2) \equiv \frac{1}{9} \int_0^1 du dv\,
\phi_{M_1}(v)\, \phi_{M_2}(u) \,\frac{r_k(u,v)}{\bar v}.
\ee
Integrating the kernels (\ref{kernel1}), (\ref{kernel2}),  
and truncating the Gegenbauer expansions after $n=2$, 
we obtain 
\begin{eqnarray}
        R_1 &=&  C_F \bigg[\! - \hf\, \ell^2+ \hf\, \ell + 
                  \frac{9}{2} - \frac{3 \pi^2}{4}
                  + 2 i \pi \bigg] 
\nonumber\\
&&   \hspace{-0.15cm}   +\, \bigg(\!C_F- \frac{C_A}{2}\bigg)
                        \bigg[ \bigg(2 
                + \frac{2 \pi^2}{3}\bigg) \ell
                -\frac{74}{5} - 2 \pi^2 + \frac{32}{5} \zeta(3)
                - i \pi \bigg(1 + \frac{2 \pi^2}{5}\bigg) \bigg]
\nonumber\\
&& \hspace{-0.5cm}  +\,a^{M_2}_1 \, \Bigg\{
        C_F \bigg[\! - \hf\, \ell^2 - \frac{13}{6}\, \ell
          + \frac{175}{18} + \frac{7 \pi^2}{12}
          + i\pi \frac{14}{3} \bigg]
\nonumber\\
&& \hspace{-0.15cm}      +\, \bigg(\!C_F- \frac{C_A}{2}\bigg) \bigg[
          \bigg( 2 + \frac{2 \pi^2}{3}\bigg) \ell
            -\frac{442}{15} - \frac{2 \pi^2}{3} + \frac{32}{5} \zeta(3)
            - i\pi \bigg(3 + \frac{2 \pi^2}{5} \bigg)
        \bigg] \Bigg\}
\nonumber\\
&& \hspace{-0.5cm}  +\,a^{M_2}_2 \, \Bigg\{
        C_F\bigg[\! - \hf\, \ell^2 - \frac{11}{3}\, \ell
          + \frac{2741}{72} - \frac{3 \pi^2}{4}
          + i\pi \frac{37}{6} \bigg]
\nonumber\\
&& \hspace{-0.15cm}      +\, \bigg(\!C_F- \frac{C_A}{2}\bigg) \bigg[
          \bigg( 2 + \frac{2 \pi^2}{3}\bigg) \ell
            -\frac{5717}{140} + \frac{2 \pi^2}{9} + \frac{164}{35} \zeta(3)
            - i\pi \bigg( 1 + \frac{62 \pi^2}{105} \bigg)
        \bigg] \Bigg\}
\nonumber\\
&& \hspace{-0.5cm}  +\,a^{M_1}_1 \, \Bigg\{
        C_F\bigg[\! - \hf\, \ell^2+ \hf\, \ell
                - \frac{67}{30} + \frac{5 \pi^2}{36} - \frac{24}{5} \zeta(3)
                + i\pi \bigg( \frac{70}{9} - \frac{8 \pi^2}{15} \bigg) \bigg]
\nonumber\\
&& \hspace{-0.15cm}      +\, \bigg(\!C_F- \frac{C_A}{2}\bigg) \bigg[
                \bigg(\!-12 + 2 \pi^2\bigg) \ell
                + \frac{283}{15} - \frac{16 \pi^2}{3} + \frac{72}{5} \zeta(3)
                + i\pi \bigg(\frac{31}{3} - \frac{26 \pi^2}{15} \bigg)
        \bigg] \Bigg\}
\nonumber\\
&& \hspace{-0.5cm}  +\,a^{M_1}_2 \, \Bigg\{
        C_F\bigg[\! - \hf\, \ell^2+ \hf\, \ell
          + \frac{267}{14} - \frac{127 \pi^2}{36} + \frac{24}{7} \zeta(3)
          + i\pi \bigg(\!-\frac{19}{18} + \frac{8 \pi^2}{21} \bigg) \bigg]
\nonumber\\
&& \hspace{-0.15cm}      +\, \bigg(\!C_F- \frac{C_A}{2}\bigg) \bigg[
          \bigg(\!-\frac{94}{3} + 4 \pi^2 \bigg) \ell
            + \frac{9713}{140} - \frac{103 \pi^2}{9} + \frac{804}{35} \zeta(3)
            + i\pi \bigg( \frac{301}{9} - \frac{144 \pi^2}{35} \bigg)
        \bigg] \Bigg\}
\nonumber\\
&& \hspace{-0.5cm}  +\,a^{M_1}_1 a^{M_2}_1 \, \Bigg\{
        C_F\bigg[\! - \hf\, \ell^2 - \frac{13}{6}\, \ell
          + \frac{20077}{210} - \frac{89 \pi^2}{12} - \frac{216}{35} \zeta(3)
          + i\pi \bigg(\frac{34}{3} - \frac{24 \pi^2}{35} \bigg) \bigg]
\nonumber\\
&& \hspace{-0.15cm}      + \, \bigg(\!C_F- \frac{C_A}{2}\bigg) \bigg[
          \bigg(\!-12 + 2 \pi^2\bigg) \ell
          + \frac{68717}{315} - \frac{76 \pi^2}{3} + \frac{456}{35} \zeta(3)
          + i\pi \bigg( \frac{29}{3} - \frac{66 \pi^2}{35} \bigg)
        \bigg] \Bigg\}
\nonumber\\
&& \hspace{-0.5cm}  +\,a^{M_1}_1 a^{M_2}_2 \, \Bigg\{
        C_F\bigg[\! - \hf\, \ell^2  - \frac{11}{3}\, \ell
          -\frac{56293}{420} + \frac{205 \pi^2}{12} - \frac{108}{35} \zeta(3)
          + i\pi \bigg(\frac{115}{12} - \frac{12 \pi^2}{35} \bigg) \bigg]
\nonumber\\
&& \hspace{-0.15cm}      + \, \bigg(\!C_F- \frac{C_A}{2}\bigg) \bigg[
          \bigg(\!-12 + 2 \pi^2\bigg) \ell
          + \frac{855199}{2520} - \frac{227 \pi^2}{6} + \frac{384}{35} \zeta(3)
          + i\pi \bigg( \frac{139}{12} - \frac{74 \pi^2}{35} \bigg)
        \bigg] \Bigg\}
\nonumber\\
&& \hspace{-0.5cm}  +\,a^{M_1}_2 a^{M_2}_1 \, \Bigg\{
        C_F\bigg[\! - \hf\, \ell^2 - \frac{13}{6}\, \ell
          -\frac{64199}{168} + \frac{457 \pi^2}{12} + \frac{108}{7} \zeta(3)
          + i\pi \bigg(\! -\frac{149}{12} + \frac{12 \pi^2}{7} \bigg) \bigg]
\nonumber\\
&& \hspace{-0.15cm}      + \, \bigg(\!C_F- \frac{C_A}{2}\bigg) \bigg[
          \bigg(\!-\frac{94}{3} + 4 \pi^2 \bigg) \ell
          -\frac{111623}{210} + \frac{148 \pi^2}{3} + \frac{804}{35} \zeta(3)
          \nonumber\\
&& \hspace*{3cm}
          +\, i\pi \bigg( \frac{187}{6} - \frac{144 \pi^2}{35} \bigg)
        \bigg] \Bigg\}
\nonumber\\ 
&& \hspace{-0.5cm}  +\,a^{M_1}_2 a^{M_2}_2 \, \Bigg\{
        C_F\bigg[\! - \hf\, \ell^2 - \frac{11}{3}\, \ell
          + \frac{711031}{504} - \frac{1709 \pi^2}{12} + \frac{144}{7} \zeta(3)
          + i\pi \bigg(\! -\frac{49}{3} + \frac{16 \pi^2}{7} \bigg) \bigg]
\nonumber\\
&& \hspace{-0.15cm}      + \, \bigg(\!C_F- \frac{C_A}{2}\bigg) \bigg[
          \bigg(\!-\frac{94}{3} + 4 \pi^2 \bigg) \ell
         -\frac{3348089}{3780} + \frac{503 \pi^2}{6} + \frac{1044}{35} \zeta(3)
          \nonumber\\
&& \hspace*{3cm}
+ \,i\pi \bigg( \frac{829}{36} - \frac{352 \pi^2}{105} \bigg)
        \bigg] \Bigg\},
\end{eqnarray}
and
\begin{eqnarray}
  R_2  &=& 3\,\ell -\frac{163}{20} + \frac{\pi^2}{3} 
                - \frac{14}{5} \zeta(3)
                + i\pi \bigg(\! -3 + \frac{2 \pi^2}{15} \bigg)
\nonumber \\
&& \hspace{-0.5cm} +\,\, a^{M_2}_1 \bigg\{ 3\,\ell -\frac{353}{20} 
   + \frac{8 \pi^2}{3}
                - \frac{54}{5} \zeta(3)
                + i\pi \bigg(\! -4 + \frac{2 \pi^2}{15} \bigg)
          \,\bigg\} 
\nonumber \\
&& \hspace{-0.5cm} +\,\, a^{M_2}_2 \bigg\{ 3\,\ell -\frac{3751}{140} 
   + \frac{16 \pi^2}{3}
                - \frac{828}{35} \zeta(3)
                + i\pi \bigg(\! -3 + \frac{4 \pi^2}{105} \bigg)
          \,\bigg\} 
\nonumber \\
&& \hspace{-0.5cm} +\, \,a^{M_1}_1 \bigg\{ 3\,\ell + \frac{259}{60} 
   + \frac{4 \pi^2}{9}
                - \frac{66}{5} \zeta(3)
                + i\pi \bigg(\! -\frac{7}{9} - \frac{2 \pi^2}{15}  \bigg)
          \,\bigg\} 
\nonumber \\
&& \hspace{-0.5cm} +\,\, a^{M_1}_2 \bigg\{ 3\,\ell + \frac{1297}{120} 
   + \frac{8 \pi^2}{9}
                - \frac{114}{5} \zeta(3)
                + i\pi \bigg(\! -\frac{32}{9} + \frac{2 \pi^2}{15} \bigg)
          \,\bigg\} 
\nonumber \\
&& \hspace{-0.5cm} +\,\,a^{M_1}_1 a^{M_2}_1 \bigg\{ 3\,\ell 
   + \frac{12263}{84} - \frac{31 \pi^2}{3}
                - \frac{270}{7} \zeta(3)
                + i\pi \bigg(\! -\frac{2}{3} - \frac{2 \pi^2}{7} \bigg)
          \,\bigg\} 
\nonumber \\
&& \hspace{-0.5cm} +\,\,a^{M_1}_1 a^{M_2}_2 \bigg\{ 3\,\ell 
    + \frac{47811}{140} - \frac{77 \pi^2}{3}
                - \frac{2592}{35} \zeta(3)
                + i\pi \bigg(\! -\frac{4}{3} - \frac{8 \pi^2}{35} \bigg)
          \,\bigg\} 
\nonumber \\
&& \hspace{-0.5cm} +\,\,a^{M_1}_2 a^{M_2}_1 \bigg\{ 3\,\ell 
   -\frac{60541}{240} + \frac{397 \pi^2}{12}
                - \frac{324}{5} \zeta(3)
                + i\pi \bigg(\! -\frac{281}{24} + \frac{4 \pi^2}{5} \bigg)
          \,\bigg\}
\nonumber \\
&& \hspace{-0.5cm}
        + \,\,a^{M_1}_2 a^{M_2}_2\bigg\{3\,\ell -\frac{2026519}{2160}
        + \frac{443 \pi^2}{4} - \frac{654}{5} \zeta(3)
                + i\pi \bigg(\! -\frac{1331}{72} + \frac{22 \pi^2}{15} \bigg)
          \,\bigg\}.
\end{eqnarray}
Here we defined $\ell \equiv \ln (m_b^2/\mu^2)$. The finiteness 
of $R_{1,2}$ proves factorization of spectator scattering to the
1-loop order. For $\mu=m_b$ 
($\ell=0$) the magnitude of the correction and importance of the
higher Gegenbauer moments can be seen from the numerical 
expressions 
\begin{eqnarray}
        R_1 &=& 0.6047  + 10.9683\,i
+(25.36 + 23.19\,i) \,a^{M_2}_1
+(46.39 + 29.41\,i) \,a^{M_2}_2
\nonumber \\ && \qquad
+(-6.10 + 14.08\,i) \,a^{M_1}_1
+(-12.84 + 15.08\,i) \,a^{M_1}_2
\nonumber \\ && \qquad
+(22.69 + 23.81\,i) \,a^{M_1}_1 a^{M_2}_1
+(44.63 + 30.83\,i) \,a^{M_1}_1 a^{M_2}_2
\nonumber \\ && \qquad
+(19.21 + 23.80\,i) \,a^{M_1}_2 a^{M_2}_1
+(43.62 + 31.35\,i) \,a^{M_1}_2 a^{M_2}_2,
\\[0.2cm]
R_2 &=& -8.2259 - 5.2906\,i
-(4.31 + 8.43 \,i) \,a^{M_2}_1
-(2.59 + 8.24\,i) \,a^{M_2}_2
\nonumber \\* && \qquad
-(7.16 + 6.58\,i) \,a^{M_1}_1
-(7.83 + 7.04\,i) \,a^{M_1}_2
\nonumber \\* && \qquad
-(2.36 + 10.95\,i) \,a^{M_1}_1 a^{M_2}_1
-(0.83 + 11.28\,i) \,a^{M_1}_1 a^{M_2}_2
\nonumber \\* && \qquad
-(3.63 + 11.98\,i) \,a^{M_1}_2 a^{M_2}_1
-(2.37 + 12.60\,i) \,a^{M_1}_2 a^{M_2}_2.
\end{eqnarray}
For phenomenology the most important 1-loop spectator-scattering 
correction is the term 
$C_1 R_1$ involving the large Wilson coefficient $C_1$ in the 
expression for $\alpha_2$. From the above expression for $R_1$ 
we learn that this contribution has a large imaginary part, 
while the real part seems to be accidentally small. The higher 
Gegenbauer moments have relatively large coefficients. Roughly, 
the magnitude of the correction to $h_1(M_1 M_2)$ is 
$14\,\alpha_s/(4\pi)\approx \alpha_s\approx 0.3$, i.e.~a 30\% correction 
relative to the tree approximation. A detailed 
numerical analysis of the tree amplitudes will be performed below. 

\subsection{Scale issues} 
\label{sec:scaleresum}

Up to this point we did not make explicit the scale dependences 
of coupling constants and parameters. The tree amplitudes $\alpha_{1,2}$ 
themselves are scale-independent, but the Wilson coefficients $C_i$, 
strong coupling $\alpha_s$, static $B$ decay constant $\hat{f}_B$, 
light-cone distribution amplitudes of the light mesons (hence the 
Gegenbauer moments $a_i^M$), as well as the $B$ meson distribution 
amplitude moments $\lambda_B$, $\langle L^n\rangle$ are all 
scale-dependent. Expressing all these quantities at the scale 
$\mu$ equal to the one that appears explicitly in the 1-loop 
result for the hard-scattering kernels is a legitimate choice.  
However, with any single scale one or another kernel will contain 
parametrically large logarithms $\ln m_b/\Lambda_{\rm QCD}$.

For the following discussion we assume that the hadronic quantities 
$a_i^M$, $\lambda_B$ as well as the logarithmic moments 
\be 
\sigma_n(\mu_0) \equiv \lambda_B(\mu_0) \int_0^\infty \frac{d\omega}{\omega} 
\,\phi_{B+}(\omega)\,\ln^n \frac{\mu_0}{\omega}
\ee
related to $\langle L^n\rangle$ are given at a reference scale $\mu_0$
of order of the hard-collinear scale $(m_b\Lambda_{\rm QCD})^{1/2}$. 
The first line of (\ref{alphapar}), which corresponds to the 
form factor term in the factorization formula (\ref{eq:qcdfact}) 
or (\ref{scet2fact}) is easy, since it contains only a hard 
correction. There are no large logarithms when the Wilson coefficients
and vertex kernels $V_i(M_2)$ are evaluated at a common scale 
$\mu_b$ of order $m_b$. Since the tree approximation is independent 
of the Gegenbauer moments, the Gegenbauer moments in 
$V_i(M_2)$ are evolved from $\mu_0$ to $\mu_b$ in the leading-logarithmic 
(LL) approximation with the 1-loop anomalous dimension matrix. 

The spectator scattering term is more involved. In order to resum 
large logarithms one should perform the substitution
\bea
&&C(\mu)\,H^\mathrm{II}(\mu) \ast J_\parallel(\mu) 
\ast [\hat f_{B}\phi_{B+}](\mu) \ast \phi_{M_1}(\mu)\ast
\phi_{M_2}(\mu)
\nonumber\\
&&\hspace*{0cm}
\Longrightarrow\quad 
C(\mu_b)\,H^\mathrm{II}(\mu_b) \ast U(\mu_b, \mu_\hc) 
\ast J_\parallel(\mu_\hc) \ast 
[\hat f_{B}\phi_{B+}](\mu_\hc) \ast \phi_{M_1}(\mu_\hc)\ast
\phi_{M_2}(\mu_\hc) \qquad,
\eea
where $\mu_\hc \sim (m_b \Lambda_{\rm QCD})^{1/2}$ is of order of 
the hard-collinear scale, and $U(\mu_b, \mu_\hc)$ is the evolution 
function for the SCET${}_{\rm I}$ operator $O^{\rm II}$. While in the 
first line either $H^\mathrm{II}(\mu)$ or $J_\parallel(\mu)$ contains 
large logarithms, neither $H^\mathrm{II}(\mu_b)$ nor 
$J_\parallel(\mu_\hc)$ does. The evolution function for  $O^{\rm II}$ 
factorizes into $U_{\rm BL}$ related to the Brodsky-Lepage kernel in the
collinear-2 sector and the evolution function $U_\parallel$ for a 
B1-type current. Since $U_{\rm BL}(\mu_b, \mu_\hc)\ast 
\phi_{M_2}(\mu_\hc)=\phi_{M_2}(\mu_b)$, we can rewrite the 
previous expression as 
\be 
C(\mu_b)\,H^\mathrm{II}(\mu_b) \ast U_\parallel(\mu_b, \mu_\hc) 
\ast J_\parallel(\mu_\hc) \ast 
[\hat f_{B}\phi_{B+}](\mu_\hc) \ast \phi_{M_1}(\mu_\hc)\ast
\phi_{M_2}(\mu_b).
\label{resum}
\ee
In this expression the Gegenbauer moments of $M_2$ must be 
evolved from $\mu_0$ to $\mu_b$ in the next-to-leading-logarithmic (NLL) 
approximation using the 2-loop anomalous dimension matrix, since the 
Gegenbauer moments appear already in the tree approximation to 
spectator scattering. We have implemented the 2-loop evolution based on the 
results of \cite{Floratos:1977au}. The evolution preserves 
the truncation of the Gegenbauer expansion due to the triangular
structure of the anomalous dimension matrix. Up to the 
second Gegenbauer moment the evolution is obtained from  
\be
\frac{d}{d\ln\mu} a^M_k(\mu) = \bigg[\gamma_{kl}^{(0)} 
\frac{\alpha_s}{4\pi} + \gamma_{kl}^{(1)} 
\frac{\alpha_s^2}{(4\pi)^2}+\ldots\bigg]\,a_l^M(\mu)
\ee
with (putting $n_f=4$)
\be
\gamma_{kl}^{(0)} = 
\left(\begin{array}{ccc} 
0 & 0 & 0\\
0 &-\frac{64}{9} & 0\\
0 & 0 &-\frac{100}{9}
\end{array}\right),
\qquad
\gamma_{kl}^{(1)} = 
\left(\begin{array}{ccc}
0 & 0 & 0\\
0 &-\frac{17344}{243} & 0\\
-\frac{100}{9} & 0 &-\frac{24490}{243}
\end{array}\right).
\ee
The Gegenbauer moments and $B$ meson 
distribution amplitude moment $\lambda_B$ in $J_\parallel(\mu_\hc) \ast 
[\hat f_{B}\phi_{B+}](\mu_\hc) \ast \phi_{M_1}(\mu_\hc)$ can be 
obtained from the input values at $\mu_0$ by a fixed-order 
1-loop relation, because no large logarithms appear when these factors 
are evaluated at $\mu_\hc$. $\hat f_{B}$ is obtained from the physical 
decay constant $f_B$ by a HQET conversion factor using the 2-loop 
approximation for the anomalous dimension and 
the 1-loop matching coefficient, 
since the matching to $f_B$ is done at the large scale $\mu_b$. To 
complete the evaluation of (\ref{resum}) 
one would now require the 2-loop anomalous dimension kernel 
for the B1-type current to evaluate $U_\parallel(\mu_b, \mu_\hc)$  
in the next-to-leading-logarithmic approximation. Only the 
1-loop anomalous dimension is available (first paper of 
\cite{Hill:2004if}, \cite{Beneke:2005gs}). We therefore implement 
an approximate procedure analogous to that in 
\cite{Beneke:2005gs} for heavy-to-light form factors. We evaluate 
$H^\mathrm{II}(\mu)$ at the hard-collinear scale $\mu_\hc$ 
except for the terms involving logarithms related to the
scale dependence of the Wilson coefficient $C$ and of $\phi_{M_2}$, 
which remain at $\mu_b$. To this expression we add the series 
of all leading logarithms summed to all orders in perturbation theory
omitting the terms already included in 
$H^\mathrm{II}(\mu_\hc)$. The structure of these terms is identical 
to eq. (117) of \cite{Beneke:2005gs}. 

Finally, we match $\alpha_s$ from a 5-flavour to a 4-flavour theory 
at the scale $\mu_b$. Quantities evaluated at $\mu_\hc$ are computed 
in the 4-flavour theory, quantities at $\mu_b$ in the 5-flavour theory.

\section{Numerical analysis of $\alpha_1$ and $\alpha_2$ and 
application to the $B\to \pi \pi$ system}
\label{sec:pipi}

\subsection{Input parameters}

For our numerical study of the tree amplitudes $\alpha_1$ and
$\alpha_2$ we employ the input parameters shown in 
Table~\ref{tab:inputs}. With respect to \cite{Beneke:2003zv} 
we have updated the value of $|V_{cb}|$, and reduced the 
error estimate of $a_2^\pi$. The heavy-quark masses $m_b$, $m_c$ are now 
interpreted as pole masses with $m_b=4.8\,$GeV. The list is extended 
by the logarithmic moments of the $B$ meson distribution 
amplitudes, for which we use values in the ranges obtained 
from QCD sum rules or models for the shape of 
the distribution amplitude~\cite{Braun:2003wx}. The 
hard-collinear scale $\mu_\hc$ and the hard matching scale $\mu_b$ 
are varied independently within the given ranges.

\renewcommand{\arraystretch}{1.4}
\begin{table}
\begin{center}
\begin{tabular}{|cc|cc|}\hline
Parameter & Value/Range & Parameter & Value/Range \\
\hline
$\Lambda^{\overline{\mathrm{MS}}(5)}$ & 0.225 & 
$\mu_b$ & $4.8^{+4.8}_{-2.4}$ \\
$m_c$ & $1.3\pm 0.2$  & 
$\mu_\mathrm{hc}$ & $1.5\pm 0.6$ \\
$m_s$(2 GeV) & $0.09\pm 0.02$ & 
$f_{B_d}$ & $0.20 \pm 0.03$  \\
$m_b$ & 4.8 & 
$f^{B\pi}_+(0)$ &  $0.28 \pm 0.05$  \\
$\bar m_b(\bar m_b)$ & 4.2 & 
$\lambda_B$(1 GeV) & $0.35 \pm 0.15$  \\
$|V_{cb}|$ & $0.0415\pm 0.0010$  & 
$\sigma_1$(1 GeV) &  $1.5 \pm 1$\\
$|V_{ub}/V_{cb}|$ &  $0.09\pm 0.02$  & 
$\sigma_2$(1 GeV) &  $3 \pm 2$\\
$\gamma$ & $(70 \pm 20)^\circ$ & 
$a^\pi_2$(2 GeV) &  $0.1 \pm 0.2$ \\
\hline
\end{tabular}
\end{center}
\caption{List of input parameters. Dimensionful parameters are given
in units of 1 GeV. \label{tab:inputs}}
\end{table}

\subsection{Tree amplitudes $\alpha_1$ and $\alpha_2$}

Numerically, we obtain the tree amplitudes\footnote{The 
following numbers differ from \cite{Beneke:2005yv}, 
since in \cite{Beneke:2005yv} the hard 1-loop spectator-scattering 
correction has been added to the program used 
in~\cite{Beneke:2003zv} to allow for a direct comparison with 
the scenarios defined there. For the present numerical 
evaluation the program has been substantially changed in order 
to implement the various scale dependences as described in 
Section~\ref{sec:scaleresum}. In particular, the Wilson coefficients 
$C_i$ are now evaluated at $\mu_b$, and always in the NLL approximation.}
\begin{eqnarray}
\alpha_1(\pi\pi) &=& 1.015 + [0.025+0.012i\,]_{V}
- \left(\frac{r_{\rm sp}}{0.412}\right) 
\Big\{[0.014+0.024\delta_{a_2}]_{\rm LO} 
\nonumber\\ 
&& + \, [0.024+0.044\delta_{a_2}+ 0.020 i+0.050\delta_{a_2}\,]_{\rm
  NLO} + [0.009+0.007\delta_{a_2}]_{\rm tw3} \Big\}
\nonumber\\
&=& 0.992^{+0.029}_{-0.054}+(-0.007^{+0.018}_{-0.035})\,i,
\label{al1num}\\
\alpha_2(\pi\pi) &=& 0.184 - [0.152+0.077i\,]_{V}
+ \left(\frac{r_{\rm sp}}{0.412}\right) 
\Big\{[0.088+0.150\delta_{a_2}]_{\rm LO} 
\nonumber\\ 
&& + \, [0.029+0.130\delta_{a_2}+ 0.034 i+0.100\delta_{a_2}\,]_{\rm
  NLO} + [0.056+0.041\delta_{a_2}]_{\rm tw3} \Big\}
\nonumber\\
&=& 0.205^{+0.171}_{-0.110}+(-0.043^{+0.083}_{-0.065})\,i.
\label{al2num}
\end{eqnarray}
In these expressions we separated the tree (first number), vertex 
correction (indexed by $V$) and the spectator-scattering correction 
(remainder). The latter is further divided into the tree ($\alpha_s$, 
indexed LO), 1-loop ($\alpha_s^2$, indexed NLO), and 
twist-3 power correction. The 1-loop correction is the sum 
of the jet function and hard correction, see (\ref{hs}). We also 
made explicit the parameter dependences that are responsible 
for the bulk of the theoretical uncertainty given in 
the last line in the expressions for $\alpha_{1,2}$. (Theoretical 
errors computed from the ranges in Table~\ref{tab:inputs} are added 
in quadrature.) The most
important such parameter is the combination 
\be
r_{\rm sp} = \frac{9 f_{\pi}\hat f_B}
{m_b f_+^{B\pi}(0)\lambda_B},
\ee
which normalizes the spectator-scattering term as can be seen 
from (\ref{alphapar}). The second most important parameter 
is the second Gegenbauer moment of the pion distribution amplitude. 
This dependence is shown (for spectator scattering only, where it is 
important) to linear order in the deviation $\delta_{a_2}=a_2^\pi(2\,
\mbox{GeV})-0.1$
from the default value. The result for the two tree amplitudes 
is shown in Figure~\ref{figalpha}, which also displays a comparison 
with the leading-order and next-to-leading order (1-loop vertex
correction, tree spectator scattering) result, as well as the 
main parameter dependences. It is evident from (\ref{al1num}), 
(\ref{al2num}) or the Figure that the 1-loop spectator-scattering
correction is significant, but not large enough to put the
perturbative approach into question. For $\alpha_1$ the 1-loop 
correction exceeds the tree spectator correction, 
because the 1-loop correction 
is multiplied by the large Wilson coefficient $C_1$. However, 
the correction is small in absolute terms. For $\alpha_2$ the
correction amounts to about $35\%$ of the tree spectator-scattering 
term. Since the imaginary part is generated only at NLO, it 
is best compared to the imaginary part of the vertex correction  
$V$. This shows that the spectator-scattering correction at order 
$\alpha_s^2$ is almost as large as the vertex correction at order 
$\alpha_s$, but comes with an opposite sign such that the phases tend 
to cancel. 

With the perturbative approach thus validated through the size of the
1-loop correction, it is evident from the Figure that the 
dominant uncertainties are due to hadronic input parameters. 
The uncertainties in $f_B$, $\lambda_B$ and $f^{B\pi}_+(0)$ 
do not exclude that $r_{\rm sp}$ is a factor of 2 larger than 
its default value 0.412. In fact, it appears that the 
data on $B\to\pi\pi$ branching fractions require such an 
enhancement~\cite{Beneke:2003zv}. Until some of these parameters 
are better determined (from theory, from other data, from fits to
non-leptonic data) there remains a large uncertainty in 
the colour-suppressed tree amplitude $\alpha_2$. The colour-allowed 
tree amplitude, however, is predicted to be close to 1 with an 
uncertainty of $10\%$ even with present parameter inaccuracies. 

\begin{figure}[t]
   \vspace{0cm}
\centerline{\includegraphics[width=12.5cm]{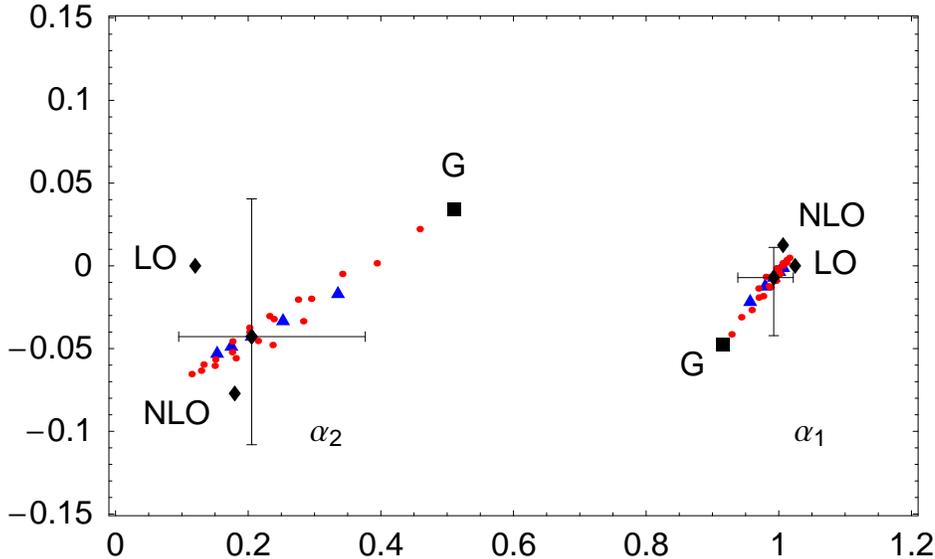}}
   \vspace*{0.2cm}
\caption{\label{figalpha} The tree amplitudes 
$\alpha_1(\pi\pi)$ and $\alpha_2(\pi\pi)$ represented in the 
complex plane. The dark (black) diamonds show the LO, 
NLO, and partial NNLO approximations. The latter 
includes the new 1-loop correction to spectator scattering and 
is shown with error bars. The dark square represents the 
parameter set `G', which provides a good description of 
the experimental data on branching fractions as discussed 
in Section~\ref{sec:brpi}. The grey (blue) triangles 
show the variation of the tree amplitudes, when $\lambda_B$ 
takes the values $0.2\,$GeV to $0.5\,$GeV in steps of 
$75\,$MeV, such that the triangles in the direction of the 
point `G' correspond to smaller values of $\lambda_B$. From each 
triangle emanates a set of grey (red) points that correspond to 
varying $a_2^\pi$ from $-0.1$ to 0.3 in steps of 0.1 for 
the given value of $\lambda_B$. Here points lying towards 
`G' correspond to larger $a_2^\pi$.}
\end{figure}

\subsection{$B\to \pi\pi$ branching fractions}
\label{sec:brpi}

We confront our new (partial) NNLO results with the experimental data 
on the three tree-dominated $B\to\pi\pi$ branching ratios. 
The $B\to \pi\pi$ amplitudes are given by
\begin{eqnarray}
   \sqrt2\,{\cal A}_{B^-\to\pi^-\pi^0}
   &=& i \,\frac{G_F}{\sqrt{2}} m_B^2 f_\pi f^{B\pi}_+(0) 
       V_{ub} V_{ud}^*\Big[\alpha_1+\alpha_2\Big], 
   \nonumber\\
   {\cal A}_{\bar B^0\to\pi^+\pi^-}
   &=&  i \,\frac{G_F}{\sqrt{2}} m_B^2 f_\pi f^{B\pi}_+(0) 
       \left\{ V_{ub} V_{ud}^* \Big[\alpha_1 + \hat \alpha_4^u \Big] +
         V_{cb} V_{cd}^* \,\hat \alpha_4^c\right\}, 
   \nonumber\\
   -\,{\cal A}_{\bar B^0\to\pi^0\pi^0}
   &=&  i \,\frac{G_F}{\sqrt{2}} m_B^2 f_\pi f^{B\pi}_+(0) 
       \left\{ V_{ub} V_{ud}^* \Big[\alpha_2 
    - \hat\alpha_4^u\Big] - 
         V_{cb} V_{cd}^*\,\hat\alpha_4^c \right\}, 
\label{pirhoampsimp}
\end{eqnarray}
not showing some smaller amplitudes that are taken into account in 
the numerical evaluation of the branching fractions below. The theoretical 
computation includes the 1-loop correction to  
spectator scattering in the tree amplitudes, $\alpha_{1,2}$, 
but not to the QCD penguin amplitudes $\hat \alpha_4^{u,c}$. 
For these (and the smaller amplitudes not shown) 
the NLO expression \cite{Beneke:1999br,Beneke:2003zv} updated to include 
the scale-dependent parameters $a_2^\pi$ and $\hat{f}_B$ in the 
LO approximation for spectator scattering is used. 

The standard input parameter set does not provide an adequate 
description of $B\to\pi\pi$ data. Rather 
the data favours a smaller value of 
$|V_{ub}| \,f^{B\pi}_+(0)$, which reduces the overall normalization 
of the amplitudes, and a significantly 
larger contribution from spectator scattering, which 
increases $\alpha_2$ (see Figure~\ref{figalpha})~\cite{Beneke:2003zv}.  
We find that the parameter choice `G' with 
\be
|V_{ub}| \,f^{B\pi}_+(0) = 8.10 \cdot 10^{-4} = 
0.775 \,\Big[|V_{ub}| \,f^{B\pi}_+(0)\Big]_{\rm def}, 
\qquad r_{\rm sp}=1.96\, [r_{\rm sp}]_{\rm def}
\label{defG}
\ee
and $a_2^\pi(2\,\mbox{GeV})=0.3$ yields a good description of data. The 
required parameter modification is most likely related to a smaller value for 
the $B\to \pi$ form factor and a smaller value of $\lambda_B$, 
but other small modifications may add up 
to the combined effect. The new parameter selection `G' is similar to 
scenario S4 defined in~\cite{Beneke:2003zv}, and falls within the 
ranges for the individual parameters specified in 
Table~\ref{tab:inputs}. With (\ref{defG}) we calculate the 
CP-averaged branching fractions 
\bea
10^6\,\mbox{Br}(B^-\to \pi^-\pi^0) &=& 
5.5^{+0.3}_{-0.3}(\mbox{CKM})^{+0.5}_{-0.4}(\mbox{hadr.})^{+0.9}_{-0.8}
(\mbox{pow.})  \qquad [5.5\pm 0.6],
\nonumber\\
10^6\,\mbox{Br}(\bar{B}^0\to \pi^+\pi^-) &=& 
5.0^{+0.8}_{-0.9}(\mbox{CKM})^{+0.3}_{-0.5}(\mbox{hadr.})^{+1.0}_{-0.5}
(\mbox{pow.})\qquad [5.0\pm 0.4],
\\
10^6\,\mbox{Br}(\bar{B}^0\to \pi^0\pi^0) &=& 
0.73^{+0.27}_{-0.24}(\mbox{CKM})^{+0.52}_{-0.21}
(\mbox{hadr.})^{+0.35}_{-0.25}(\mbox{pow.}) \qquad [1.45\pm 0.29]
\nonumber
\eea
with the experimental averages reproduced in brackets
\cite{HFAG}. The corresponding tree amplitudes  
$\alpha_1 = 0.92-0.05 i$ and $\alpha_2 =  0.51 + 0.03 i$ 
are shown by the points marked `G' in Figure ~\ref{figalpha}, 
which implies that 
the ratio of the colour-suppressed to colour-allowed amplitude 
$C/T=\alpha_2/\alpha_1 =  0.55+0.07 i$ is large.
By construction the branching fractions with charged
pions in the final state are in excellent agreement with data.   
The $B\to \pi^0\pi^0$ branching fraction is still somewhat low, but the 
theoretical uncertainty is large. In computing the theoretical 
errors we did not include here the uncertainties in $|V_{ub}|$, 
$f_+^{B\pi}(0)$, $f_B$, $\lambda_B$ and $a_2^\pi$, because these 
5 input parameters appear only through (\ref{defG}). The remaining 
uncertainties are divided into groups from $|V_{cb}|$, $\gamma$ 
(CKM); the renormalization scales $\mu_\hc$, $\mu_b$, and quark masses 
$m_c$, $m_c$ (hadr.); and $X_H$, $X_A$ (weak annihilation)  
parameterizing non-factorizable power corrections (pow.). The decays to 
the final states $\pi^+\pi^-$ and $\pi^0\pi^0$ are sensitive to 
$\gamma$ as can be seen from the CKM uncertainty. The dominant 
errors come from the hard-collinear factorization scale 
$\mu_\hc$, and from power corrections. 
We postpone a detailed assessment 
of the theoretical status after the calculation of the 1-loop 
spectator-scattering correction to the penguin amplitudes, 
which may be important for $\bar{B}^0\to \pi^0\pi^0$. We also 
expect the spectator-scattering phase in the penguin amplitudes 
to affect the direct CP asymmetries, and therefore do not 
discuss them now.\footnote{A calculation related to the penguin 
contractions of spectator-scattering  amplitudes has appeared 
recently~\cite{Li:2005wx}, but the method of computation is not exactly 
what is required for QCD $\to$ SCET${}_{\rm I}$ matching.}

\section{Conclusion}
\label{sec:conclude}

We computed the 1-loop hard spectator-scattering correction to the 
topological tree amplitudes in non-leptonic $B$ decays by 
matching the current-current operators $Q_{1,2}$ from the effective weak
interaction Hamiltonian to the relevant operators in SCET${}_{\rm I}$. 
Together with the 1-loop calculation of the hard-collinear 
jet function~\cite{Hill:2004if,Beneke:2005gs}, the 
spectator-scattering contribution is now complete at 
order $\alpha_s^2$, also providing the first and perhaps dominant 
contribution to a full NNLO computation of the decay amplitudes. 
Unless the 1-loop term is enhanced 
by a large Wilson coefficient, it is of order $(30-50)\%$ relative 
to the tree term, depending on moments of light-cone distribution
amplitudes. This yields a visible enhancement of 
the spectator-scattering amplitude without changing the qualitative
features of the previous approximation. 

The very fact that the perturbative correction can be computed, 
and that the expansion seems to be reasonably behaved, is 
significant, since it shows that a) factorization holds theoretically, 
i.e. IR singularities factorize as predicted and convolution integrals
converge, and that b) perturbative expansions of the 
spectator-scattering contribution are under control, an issue that 
has at times been a point of controversy (second paper of 
\cite{Chay:2003ju}, \cite{Beneke:2004bn}). Our
calculation therefore shows that the theoretical accuracy  
is now limited by the uncertainties of input parameters for the 
factorization formula, and encourages efforts to determine 
better key hadronic quantities such as form factors, and moments 
of light-cone distribution amplitudes.

\subsubsection*{Acknowledgements}

We wish to thank D.~Mueller for providing the 2-loop anomalous 
dimension matrix for the light-cone distribution amplitudes. 
M.B. would like to thank the INT, Seattle and KITP, Santa 
Barbara for their generous hospitality during the summer of 2004, 
when part of this work was being done.  
This work is supported in part by the 
DFG Sonder\-forschungs\-bereich/Trans\-regio~9 
``Com\-puter\-ge\-st\"utz\-te Theoretische Teilchenphysik''.

\end{document}